\documentclass[twocolumn,aps,prd,showpacs,groupedaddress,nofootinbib]{revtex4-1}
\usepackage{amsmath}
\usepackage{bm}
\usepackage{slashed}
\usepackage{epsfig}
\usepackage{amsfonts}
\usepackage{xcolor}
\usepackage{epstopdf}
\usepackage{enumitem}
\usepackage{subfig}
\usepackage[pdftex]{hyperref}

\begin{document}

\normalsize

\title{\boldmath The complete study on polarization of $\Upsilon(nS)$ hadroproduction at QCD next-to-leading order}

\author{Yu Feng$^{1,2}$}\email{Email: yfeng@ihep.ac.cn}
\author{Bin Gong$^{3,4}$}\email{Email: twain@ihep.ac.cn}
\author{Chao-Hsi Chang$^{2,4}$}\email{Email: zhangzx@itp.ac.cn}
\author{Jian-Xiong Wang$^{3,4}$}\email{Email: jxwang@ihep.ac.cn}
\affiliation{
$^{1}$ Department of Physics, College of Basic Medical Sciences, Army Medical University, Chongqing 400038, China \\
$^{2}$ CAS Key Laboratory of Theoretical Physics, Institute of Theoretical Physics, Chinese Academy of Sciences, Beijing 100190, China\\
$^{3}$ Institute of High Energy Physics, Chinese Academy of Sciences, P.O.Box 918(4), Beijing 100049, China \\
$^{4}$School of Physical Sciences, University of Chinese Academy of Sciences, Beijing 100190, China
}

\date{\today}

\begin{abstract}
  Applying the nonrelativistic quantum chromodynamics factorization formalism to the $\Upsilon(1S,2S,3S)$ hadroproduction,
  a complete analysis on the polarization parameters $\lambda_{\theta}$, $\lambda_{\theta\phi}$, $\lambda_{\phi}$ for the production
  are presented at QCD next-to-leading order.
  With the long-distance matrix elements extracted from experimental data for the production rate and polarization
  parameter $\lambda_{\theta}$ of $\Upsilon$ hadroproduction,
  our results provide a good description for the measured parameters $\lambda_{\theta\phi}$ and $\lambda_{\phi}$
  in both the helicity and the Collins-Soper frames.
  In our calculations the frame invariant parameter $\tilde{\lambda}$ is consistent in the two frames. Finally, it is pointed
  out that there are discrepancies for $\tilde{\lambda}$ between available experimental data and corresponding theoretical predictions.
\end{abstract}


\maketitle

\section{Introduction}
Heavy quarkonia are the most important laboratories to access the property of quantum chromodynamics (QCD). Due to  large masses of the heavy quarks, perturbative QCD is applicable at the parton level to the related heavy quarkonia.
However, to approach the heavy quarkonium production properly, the factorization method is crucial to involve the nonperturbative hadronization
from the quark pair to quarkonium. Non-relativistic quantum chromodynamics (NRQCD)~\cite{Bodwin:1994jh} may be the most successful effective theory in dealing with the perturbative and nonperturbative factors in the decays and productions of heavy quarkonia. With short-distance coefficients (SDC) and long-distance matrix elements (LDMEs),
NRQCD tells us how to organize the perturbative effects as double expansions in the coupling constant $\alpha_s$ and the heavy quark relative velocity $v$. In the past years, great improvements were made at the next-to-leading order (NLO) within NRQCD framework
~\cite{Campbell:2007ws,Gong:2008sn,Gong:2008ft,Gong:2010bk,Butenschoen:2010rq,Ma:2010yw,Butenschoen:2011yh,Ma:2010jj,Wang:2012is}.
The first evaluations of the QCD corrections to the color-singlet hadroproduction of $J/\psi$ and $\Upsilon$ were introduced in Refs.~\cite{Campbell:2007ws,Gong:2008sn}, where the transverse momentum $p_t$ distribution was found to be enhanced by 2-3 orders of magnitude at the high $p_t$ region and the $J/\psi$ polarization changed from transverse into longitudinal at NLO~\cite{Gong:2008sn}. Then, Gong~et.al~\cite{Gong:2008ft,Gong:2010bk} presented their $J/\psi$~\cite{Gong:2008ft} and $\Upsilon$~\cite{Gong:2010bk} production up to QCD NLO via the S-wave octet states $^1S_0^{[8]}$ and ~$^3S_1^{[8]}$. The analysis on complete NLO corrections within NRQCD framework was carried out later in references~\cite{Butenschoen:2010rq,Ma:2010yw,Butenschoen:2011yh,Ma:2010jj} to study the $J/\psi$ hadroproduction
for available experimental measurements independently.


Despite the achievements, NRQCD encounters challenges on the transverse momentum distribution of polarization for $J/\psi$ and $\Upsilon$ hadroproduction where the theoretical predictions can not describe the experimental data at QCD leading order (LO), and in some sense at NLO.
Three groups~\cite{Butenschoen:2012px,Chao:2012iv,Gong:2012ug} made great efforts to study the $J/\psi$ polarization parameter $\lambda_{\theta}$ at QCD NLO, but none of their color-octet (CO) LDMEs could reproduce the experimental measurements for the $J/\psi$ production from LHC~\cite{Aaij:2013nlm,Aaij:2014qea} under good precision in low and high $p_t$ regions simultaneously.
Later on, the $\eta_c$ hadroproduction measured by LHCb Collaboration~\cite{Aaij:2014bga}
provides another laboratory for test of NRQCD. Ref.~\cite{Butenschoen:2014dra} considered it as a challenge for NRQCD,
while Refs.~\cite{Han:2014jya, Zhang:2014ybe} found that these data are consistent with the $J/\psi$ hadroproduction data.
The complicated situation shows that, further studies and tests of NRQCD are an important task.

As regards the $\Upsilon$ production, similar progresses are achieved ~\cite{Campbell:2007ws,Gong:2008sn,Gong:2010bk,Wang:2012is} as those of $J/\psi$ production.
In comparison with the case of $J/\psi$,
it is expected  that theoretical predictions are of a better convergence in
the NRQCD expansions for $\Upsilon$ production due to heavier mass and smaller $v$.
 Consequently $\Upsilon$ production may provide an additional new place to test NRQCD.
%
The first complete NLO QCD corrections on yield and polarization of $\Upsilon(1S,2S,3S)$ were presented in Ref.~\cite{Gong:2013qka},
where the results provided a good description on the polarization of $\Upsilon(1S,2S)$ at CMS, as well as the yield data.
However, without considering the $\chi_{bJ}(3P)$ feed-down, the polarization of $\Upsilon(3S)$ has remained to be a problem.
Thereafter, two groups~\cite{Feng:2015wka,Han:2014kxa} updated the understanding of the $\Upsilon(3S)$
polarization by considering the $\chi_{bJ}(3P)$ feed-down contribution after the discovery of $\chi_{bJ}(3P)$ in the experiment measurements~\cite{Aaij:2014hla,Aaij:2014caa}. The results show a good description of the data of the $\Upsilon$ polarization.

The polarization of the $\Upsilon$ is measured through the analysis of the angular distribution of $\mu^+$ and $\mu^-$ from $\Upsilon$ decay (\cite{Beneke:1998re,Faccioli:2010kd}):

\begin{equation}\label{equ:decay}
\begin{split}
\frac{d^2\sigma}{d \cos\theta d\phi}\propto
1&+\lambda_{\theta}\cos^2\theta + \lambda_{\theta\phi}\sin(2\theta)\cos\phi \\
&+\lambda_{\phi}\sin^2\theta\cos(2\phi)
\end{split}
\end{equation}
where $\theta$ and $\phi$ refer to the polar and azimuthal angle of the $\mu^+$ in the $\Upsilon$ rest frame.
The three coefficients $\lambda_{\theta}$, $\lambda_{\theta\phi}$, $\lambda_{\phi}$,
which depend on the choice of reference system, contain the polarization information.
Although all the three coefficients provide independent information,
most theoretical studies on heavy quarkonium polarization are restricted to $\lambda_{\theta}$.
The parameter $\lambda_{\phi}$  of $J/\psi$ was studied at QCD NLO work in Ref.~\cite{Butenschoen:2012px}
with few experimental data points measured by ALICE~\cite{Abelev:2011md}.
Recently, complete predictions on the $J/\psi$ polarization were released by our group~\cite{Feng:2018ukp} and PKU group~\cite{Ma:2018qvc}, which reconciled the $\lambda_{\theta\phi}$ and $\lambda_{\phi}$ data quite well.
As for $\Upsilon$ polarization, although the three coefficients have been measured by CMS~\cite{Chatrchyan:2012woa},
the theoretical predictions on $\lambda_{\theta\phi}$ and $\lambda_{\phi}$ are still absent.
Furthermore, new measurements of the $\Upsilon$ polarization have been published by LHCb~\cite{Aaij:2017egv}.
The complete analysis of the $\Upsilon$ polarization therefore seems to be urgent,
especially for the predictions of parameters $\lambda_{\theta\phi}$ and $\lambda_{\phi}$.

In this paper, we will analyze the polarization of $\Upsilon(1S,2S,3S)$ in so-called helicity and Collins-Soper (CS) frames (see e.g. Ref.~\cite{Beneke:1998re} for more details on the polarization frames).
In addition, the value of the frame-invariant quantity $\widetilde{\lambda}$, which is defined as
\begin{equation}\label{eq:laminv}
\widetilde{\lambda}=\frac{\lambda_{\theta}+3 \lambda_{\phi}}{1 - \lambda_{\phi}},
\end{equation}
is computed and compared with experimental data.

In the following, a brief description of the framework and the LDMEs strategy is introduced in Sec.~\ref{charp:2}.
The numerical results of $\Upsilon(nS)$ polarization are presented in Sec.~\ref{charp:3}.
The summary and conclusion are given in Sec.~\ref{sec:summary}.

\section{Theory description}\label{charp:2}
\subsection{General setup}

The three polarization parameters $\lambda_{\theta}$, $\lambda_{\theta\phi}$ and $\lambda_{\phi}$ in Eq.~(\ref{equ:decay})
are defined as ~\cite{Beneke:1998re}
\begin{equation*}
\lambda_{\theta}=\frac{d\sigma_{11}-d\sigma_{00}}{d\sigma_{11}+d\sigma_{00}},
\lambda_{\theta\phi} = \frac{\sqrt{2}Re d\sigma_{10}}{d\sigma_{11}+d\sigma_{00}},
\lambda_{\phi} = \frac{d\sigma_{1,-1}}{d\sigma_{11}+d\sigma_{00}}.
\end{equation*}
Here, $d\sigma_{\lambda \lambda'}$($\lambda, \lambda'$= $0,\pm1$) is the spin density matrix elements of $\Upsilon$ hadroproduction, which depends on the choice of the polarization frames.
Following the NRQCD factorization~\cite{Bodwin:1994jh}, the spin density matrix elements can be expressed as
\begin{equation}\label{eq:ppsig}
\begin{split}
d\sigma_{\lambda \lambda'}(pp\rightarrow &HX)=\sum_{a,b,n}\int dx_1 dx_2 f_{a/p}(x_1) f_{b/p}(x_2) \\
&\times d\hat{\sigma}_{\lambda \lambda'}(ab\rightarrow (c\overline{c})_nX)\langle{\cal O}^{H}_{n}\rangle
\end{split}
\end{equation}
where $p$ is the proton, the indices $a$, $b$ run over all possible partons. $n$ denotes the color, spin and angular momentum states of
the $b\overline{b}$ intermediate states, which can be $^3S_1^{[1]}$, $^1S_0^{[8]}$, $^3S_1^{[8]}$ or $^3P_J^{[8]}$ for
$\Upsilon$, and $^3P_J^{[1]}$ or $^3S_1^{[8]}$ for $\chi_{bJ}$.
The function $f_{a/p}(x_1)$ and $f_{b/p}(x_2)$ are the parton distribution functions for the incoming protons for parton types $a$ and $b$.
The short-distance coefficients $d\hat{\sigma}$ can be calculated perturbatively and
the LDMEs $\langle{\cal O}^{H}_n \rangle$ are governed by nonperturbative QCD effects.



To include the feed-down contributions from higher excited states to $\Upsilon$,
we follow the treatment as in Ref.~\cite{Gong:2012ug},

\begin{eqnarray}
&&\begin{split}
&d\sigma_{\lambda \lambda'}^{\Upsilon(nS)}|_{\chi_{bJ}(mP)}={\cal B}[\chi_{bJ}(mP)\rightarrow \Upsilon(nS)]
\sum_{J_z,J'_z} d\sigma_{J_zJ'_z}^{\chi_{bJ}(mP)} \\
&\times  \delta_{J_z-\lambda,J'_z-\lambda'} C^{\lambda,J_z-\lambda}_{J,J_z}C^{*\lambda',J'_z-\lambda'}_{J,J'_z} (m \ge n),
\end{split} \\
\nonumber \\
&&\begin{split}
&d\sigma_{\lambda \lambda'}^{\Upsilon(nS)}|_{\Upsilon(mS)}={\cal B}[\Upsilon(nS)\rightarrow \Upsilon(mS)]d\sigma_{\lambda \lambda'}^{\Upsilon(mS)}\\
&(m > n).
\end{split}
\end{eqnarray}
where $C^{\lambda,J_z-\lambda}_{J,J_z}$ is the Clebsch-Gordan coefficient and
${\cal B}[X\rightarrow Y]$ denotes the branching ratio of $X$ decaying into $Y$.

To calculate the NRQCD prediction on the transverse momentum $p_t$ distribution of yield and polarization
for heavy quarkonium hadroproduction at QCD NLO, we use the FDCHQHP package~\cite{Wan:2014vka},
which is based on: 1) a collection of Fortran codes
for all the 87 parton level sub-processes generated by using FDC package~\cite{Wang:2004du},
2) implementation tool on job submission and numerical precision control.

However, for the soft and collinear divergence treatment involving P-wave quarkonium state, it was found recently by the authors of Ref.~\cite{Butenschoen:2019lef} that there is a mistake in the usual treatment
of tensor decomposition.
This mistakes is corrected in our FDC package \cite{Wang:2004du} and the related Fortran source is regenerated.
In fact, it is found that this mistake can only affect numerical results for a few percents.
%

\subsection{LDMEs Strategy}

The color-singlet LDMEs are estimated through wave functions at origin
\begin{eqnarray}
\nonumber\langle{\cal O}^{\Upsilon(nS)}(^{3}S^{[1]}_{1})\rangle&=&\frac{9}{2\pi}|R_{\Upsilon(nS)}(0)|^{2}, \\
 \langle{\cal O}^{\chi_{bJ}(mP)}(^{3}P^{[1]}_{J})\rangle&=&\frac{3}{4\pi}(2J+1)|R'_{\chi_{b}(mP)}(0)|^{2}.
\end{eqnarray}
where the wave functions and their derives at origin can be calculated via the potential
model~\cite{Eichten:1995ch}. For convenience, the related values are collected in Table~\ref{tab:potential}.

\begin{table}[htb]
\begin{center}
\caption{ \label{tab:potential}  Radial wave functions at the origin~\cite{Eichten:1995ch}.}
\footnotesize
\begin{tabular*}{80mm}{c@{\extracolsep{\fill}}ccc}
  \hline\hline
  $\Upsilon(nS)$ & $|R_{\Upsilon(nS)}(0)|^{2}$ & $\chi_{bJ}(mP)$ & $|R'{\chi_{b(mP)}}(0)|^{2}$ \\[0.1cm]
  \hline
  1S & 6.477 GeV$^{3}$ & 1P & 1.417 GeV$^{5}$ \\
  2S & 3.234 GeV$^{3}$ & 2P & 1.653 GeV$^{5}$ \\
  3S & 2.474 GeV$^{3}$ & 3P & 1.794 GeV$^{5}$ \\
  \hline\hline
\end{tabular*}
\end{center}
\end{table}

As regards the color-octet LDMEs, they can only be extracted from experimental data.
In our previous studies~\cite{Feng:2015wka},
three sets of LDMEs were obtained by fitting the experimental measurements on the yield, polarization
parameter $\lambda_{\theta}$ and the fractions of $\chi_{bJ}(mP)$ to $\Upsilon(nP)$ production.
Among these fitting schemes, different $\chi_{bJ}(mP)$ feed-down ratios and NRQCD factorization scales are used,
that leads to little differences in the results of production and polarization
although the differences in the values of LDMEs are sizable.
Considering the fact that the branching ratios ${\cal B}[\chi_{bJ}(3P)\rightarrow\Upsilon(1S,2S,3S)]$ are still absent from experimental data,
in this paper we take the color-octet LDMEs obtained by the default fitting scheme in Ref.~\cite{Feng:2015wka},
where a naive estimation for branching ratios that ${\cal B}[\chi_{bJ}(3P)\rightarrow\Upsilon(3S)] \simeq {\cal B}[\chi_{bJ}(2P)\rightarrow\Upsilon(2S)]$ and  ${\cal B}[\chi_{bJ}(3P)\rightarrow\Upsilon(1S,2S)]=0$ is used.
For convenience, we collect the values of color-octet LDMEs in TABLE~\ref{tab:ldme}.
The branching ratios of $\chi_{bJ}(mP)\rightarrow\Upsilon(nS)\gamma$ are taken from PDG data~\cite{Beringer:1900zz},
which can also be found in Table I of Ref.~\cite{Gong:2013qka}.

\subsection{Uncertainty Estimation}

\begin{table*}[htb]
  \caption[]{The Color-Octet LDMEs for bottomonia production(in units of 10$^{-2}$ GeV$^3$)~\cite{Feng:2015wka}.}
  \label{tab:ldme}
  \footnotesize
  \begin{tabular*}{155mm}{@{\extracolsep{\fill}}cccccc}
  \hline\hline
  state ~&~ $\langle{\cal O}^{\Upsilon(nS)}(^{1}S^{[8]}_{0})\rangle$ ~&~
  $\langle{\cal O}^{\Upsilon(nS)}(^{3}S^{[8]}_{1})\rangle$ ~&~
  $\langle{\cal O}^{\Upsilon(nS)}(^{3}P^{[8]}_{0})\rangle/m_b^2$ ~&~
  state ~&~ $\langle{\cal O}^{\chi_{b0}(mP)}(^{3}S^{[8]}_{1})\rangle$ \\ [0.1cm]
  \hline
  $\Upsilon(1S)$ ~&~ 13.6 $\pm$ 2.43 ~&~ 0.61 $\pm$ 0.24 ~&~ -0.93 $\pm$ 0.5  ~&~ $\chi_{b0}(1P)$ ~&~ 0.94 $\pm$ 0.06\\
  $\Upsilon(2S)$ ~&~ 0.62 $\pm$ 1.98 ~&~ 2.22 $\pm$ 0.24 ~&~ 0.13\footnote{There is a typo in Ref.~\cite{Feng:2015wka}}
  $\pm$ 0.43 ~&~ $\chi_{b0}(2P)$ ~&~ 1.09 $\pm$ 0.14\\
  $\Upsilon(3S)$ ~&~ 1.45 $\pm$ 1.16 ~&~ 1.32 $\pm$ 0.20 ~&~ -0.27 $\pm$ 0.25 ~&~ $\chi_{b0}(3P)$ ~&~ 0.69 $\pm$ 0.14\\
  \hline\hline
  \end{tabular*}
\end{table*}

\begin{table*}[htb]
  \caption[]{The rotated LDMEs for direct $\Upsilon(1S,2S,3S)$ production(in units of 10$^{-2}$ GeV$^3$). }
  \label{tab:LDMEs-rotate}
  \footnotesize
  \begin{tabular*}{140mm}{@{\extracolsep{\fill}}cccc}
  \hline\hline
  state ~&~ $\Lambda^{\Upsilon(nS)}_1$ ~&~ $\Lambda^{\Upsilon(nS)}_2 $ ~&~ $\Lambda^{\Upsilon(nS)}_3$  \\ [0.1cm]
  \hline
  $\Upsilon(1S)$ ~&~ 13.4 $\pm$ 2.45 ~&~ 1.12 $\pm$ 0.13 ~&~ 2.34 $\pm$ 0.07  \\
  $\Upsilon(2S)$ ~&~ 0.38 $\pm$ 1.99 ~&~ -1.43 $\pm$ 0.11 ~&~ 1.77 $\pm$ 0.05 \\
  $\Upsilon(3S)$ ~&~ 1.35 $\pm$ 0.00 ~&~ -1.14 $\pm$ 0.07 ~&~ 0.89 $\pm$ 0.03 \\
  \hline\hline
  \end{tabular*}
\end{table*}

Only the uncertainties from LDMEs are considered in this work. 
To express the uncertainty from CO LDMEs properly, a covariance-matrix method~\cite{Ma:2010jj,Gong:2012ug} is performed,
where we fix the CO LDMEs of $\chi_{bJ}(mP)$ and rotate
$\langle{\cal O}^{\Upsilon(nS)}(^{1}S^{[8]}_{0})\rangle$,  $\langle{\cal O}^{\Upsilon(nS)}(^{3}S^{[8]}_{1})\rangle$ and $\langle{\cal O}^{\Upsilon(nS)}(^{3}P^{[8]}_{0})\rangle/m_b^2$
(in TABLE.~\ref{tab:ldme}), the CO LDMEs of $\Upsilon(1S,2S,3S)$.
To illustrate the strategy in detail,
we denote the three direct LDMEs in a convenient way 
\begin{equation}
\begin{split}
&\mathcal{O}^{\Upsilon(nS)}\equiv \\
&\left(\langle{\cal O}^{\Upsilon(nS)}(^{1}S^{[8]}_{0})\rangle,
\langle{\cal O}^{\Upsilon(nS)}(^{3}S^{[8]}_{1})\rangle,
\frac{\langle{\cal O}^{\Upsilon(nS)}(^{3}P^{[8]}_{0})\rangle}{m_b^2}
\right)
\end{split}
\end{equation}
The rotation matrix $V_{\Upsilon(nS)}$ as discussed in Ref.~\cite{Ma:2010jj} is used to make the fitting variables independent.
We introduce variables $\Lambda^{\Upsilon(nS)}\equiv\mathcal{O}^{\Upsilon(nS)}V_{\Upsilon(nS)}$
for $\Upsilon(1S)$, $\Upsilon(2S)$ and $\Upsilon(3S)$, respectively, which are  obtained with only independent error for each $\Lambda_i$ in the fit.
Then the differential cross section $d\sigma$ is obtained with
\begin{equation}\label{eq:rotate}
  d\sigma=\sum \mathcal{O}_i d\hat{\sigma}_i=\sum \mathcal{O}V V^{-1}d\hat{\sigma}=\sum \Lambda V^{-1}d\hat{\sigma}
\end{equation}
where $d\hat{\sigma}_i$ are the corresponding short-distance coefficients
and $\Upsilon(nS)$ in the denotation has been suppressed for convenience.
The values of $\Lambda^{\Upsilon(nS)}_i$ are collected in TABLE.~\ref{tab:LDMEs-rotate},
and corresponding rotation matrix are:

\begin{eqnarray}\label{eq:rotateV}
&V_{\Upsilon(1S)}=
\left(
  \begin{array}{ccc}
    0.974 & 0.162 & 0.158 \\
    -0.079 & -0.413 & 0.907 \\
    -0.212 & 0.896 & 0.389 \\
  \end{array}
\right),\\
&V_{\Upsilon(2S)}=
\left(
  \begin{array}{ccc}
    0.974 & 0.0837 & 0.210 \\
    -0.0895 & -0.710 & 0.698 \\
    -0.208 & 0.699 & 0.685 \\
  \end{array}
\right),\\
&V_{\Upsilon(3S)}=
\left(
  \begin{array}{ccc}
    0.975 & 0.0498 & 0.215 \\
    -0.0908 & -0.797 & 0.597 \\
    -0.201 & 0.602 & 0.773 \\
  \end{array}
\right).
\end{eqnarray}

Then the uncertainties are obtained from LDMEs through
\begin{equation}\label{eq:err}
  \Delta f(\Lambda_1, \Lambda_2, \Lambda_3,\cdots) = \left[\sum_i\left(\frac{\partial f(\Lambda_i)}{\partial \Lambda_i} \Delta \Lambda_i\right)^2 +\cdots\right]^{\frac{1}{2}}
\end{equation}
where $f$ is a physical observable,
which can be any one of
the polarization parameters $\lambda_{\theta}$, $\lambda_{\theta\phi}$,  $\lambda_{\phi}$ and $\tilde{\lambda}$ in this paper.
$\Lambda_i$ denote the rotated LDMEs in TABLE~\ref{tab:LDMEs-rotate}.
The variables with $\Delta$ are the corresponding uncertainties,
and ``$\cdots$'' denotes the uncertainties from feed-down contributions.

\section{Numerical results}\label{charp:3}

In the numerical calculation, the CTEQ6M parton distribution functions~\cite{Pumplin:2002vw} and corresponding
two-loop QCD coupling constants $\alpha_s$ are used.
We adopt an approximation $m_b=M_H/2$ for the b-quark mass,
where the masses of the relevant bottomonia are taken from PDG~\cite{Beringer:1900zz}:
$M_{\Upsilon(nS)}=$ 9.5, 10.023, 10.355 GeV for $n=$ 1, 2, 3
and $M_{\chi_{bJ}(mP)}=$ 9.9, 10.252, 10.512 GeV for $m=$ 1, 2, 3, respectively.


The factorization, renormalization and NRQCD scales are chosen as $\mu_f$ = $\mu_r$ = $\sqrt{4m^2_b+p^2_t}$
and $\mu_{\Lambda}$ = $m_b v \approx$ 1.5 GeV, respectively.
A shift $p^H_t\approx p^{H'}_t \times (M_H/M_{H'})$ is used while considering the kinematics effect
in the feed-down from higher excited states.




\subsection{The polarization relating to CMS measurements}
\begin{figure*}[htb]
  \centering
  \includegraphics[width=5.0cm]{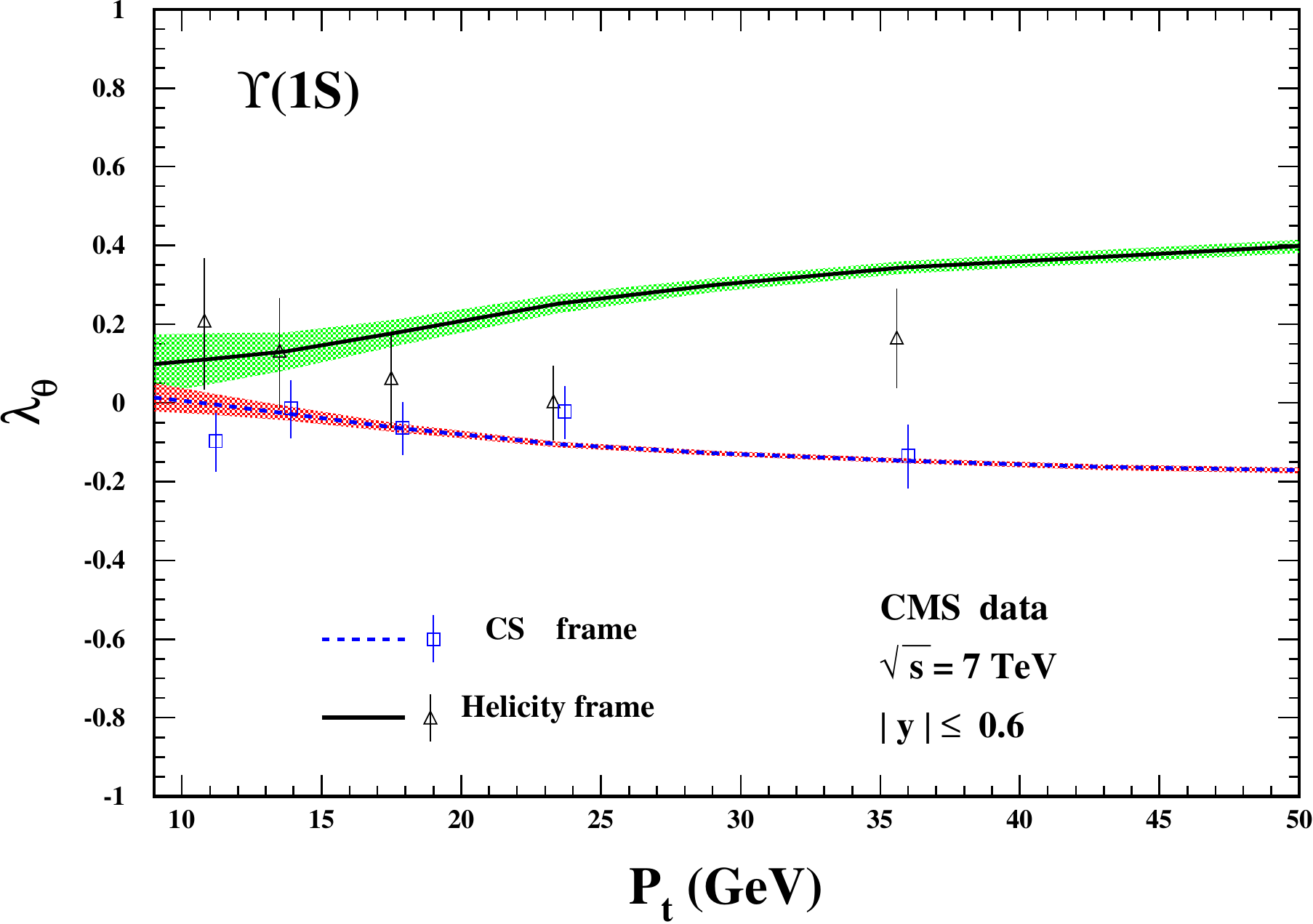}
  \includegraphics[width=5.0cm]{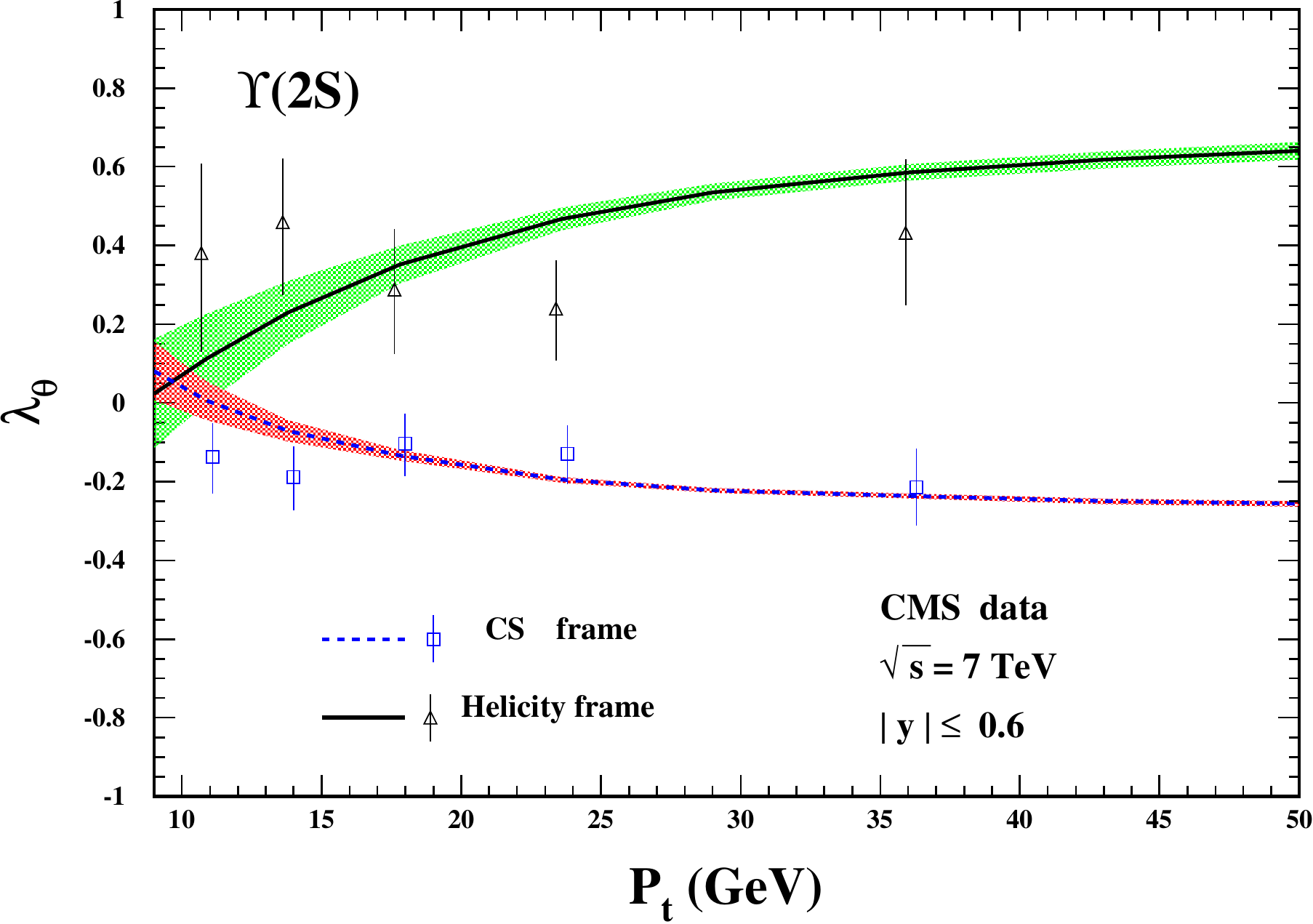}
  \includegraphics[width=5.0cm]{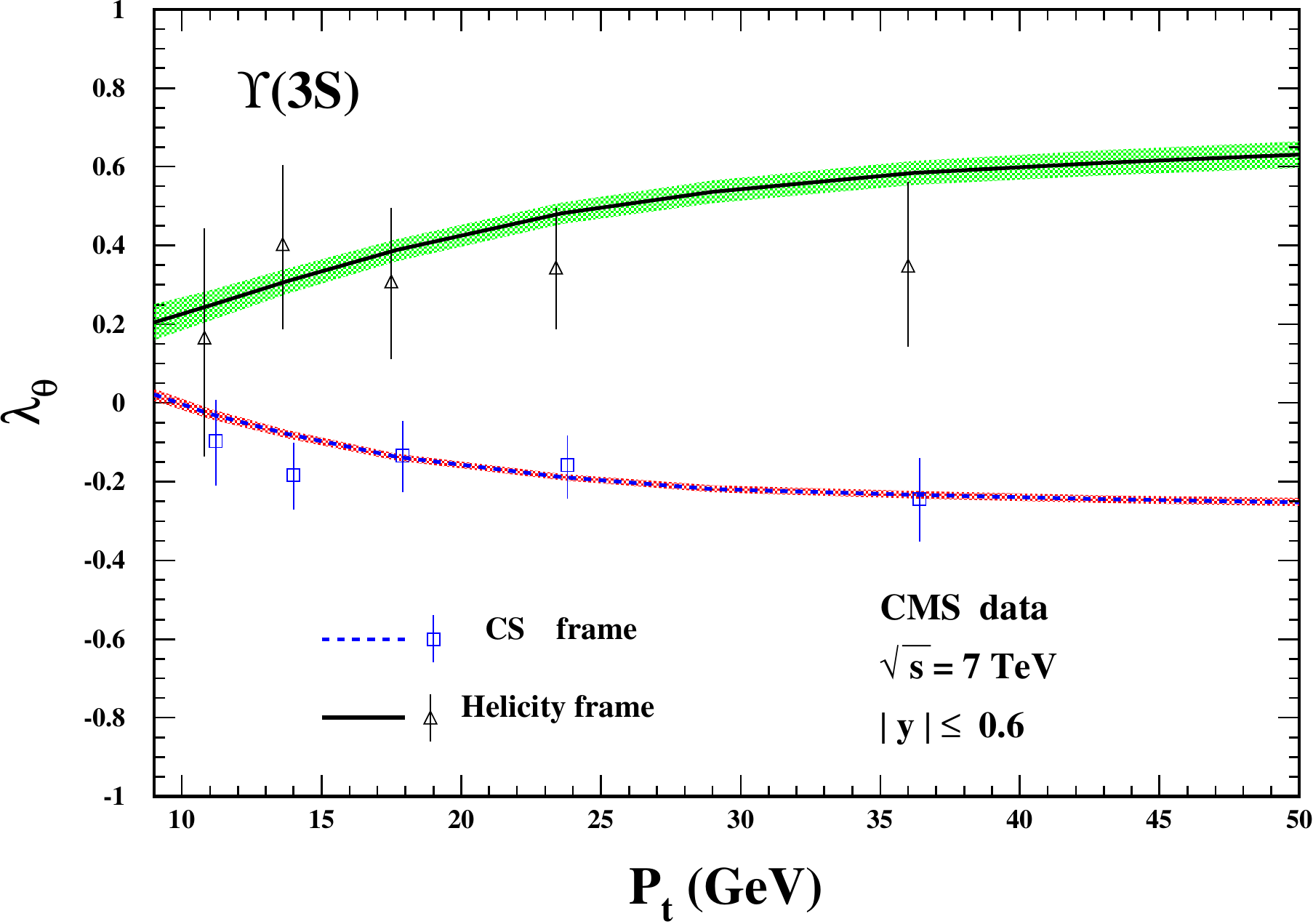} \\
  \includegraphics[width=5.0cm]{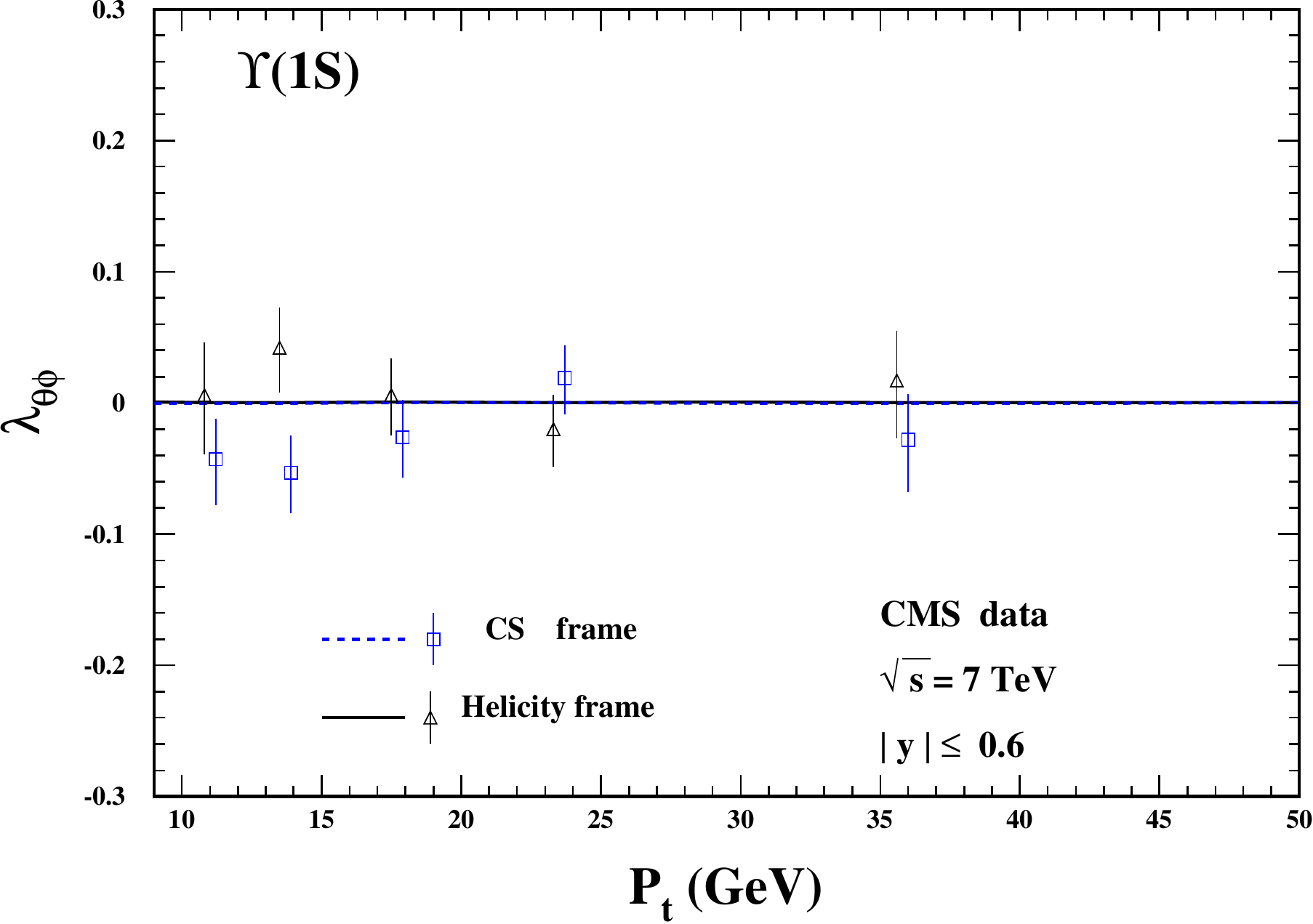}
  \includegraphics[width=5.0cm]{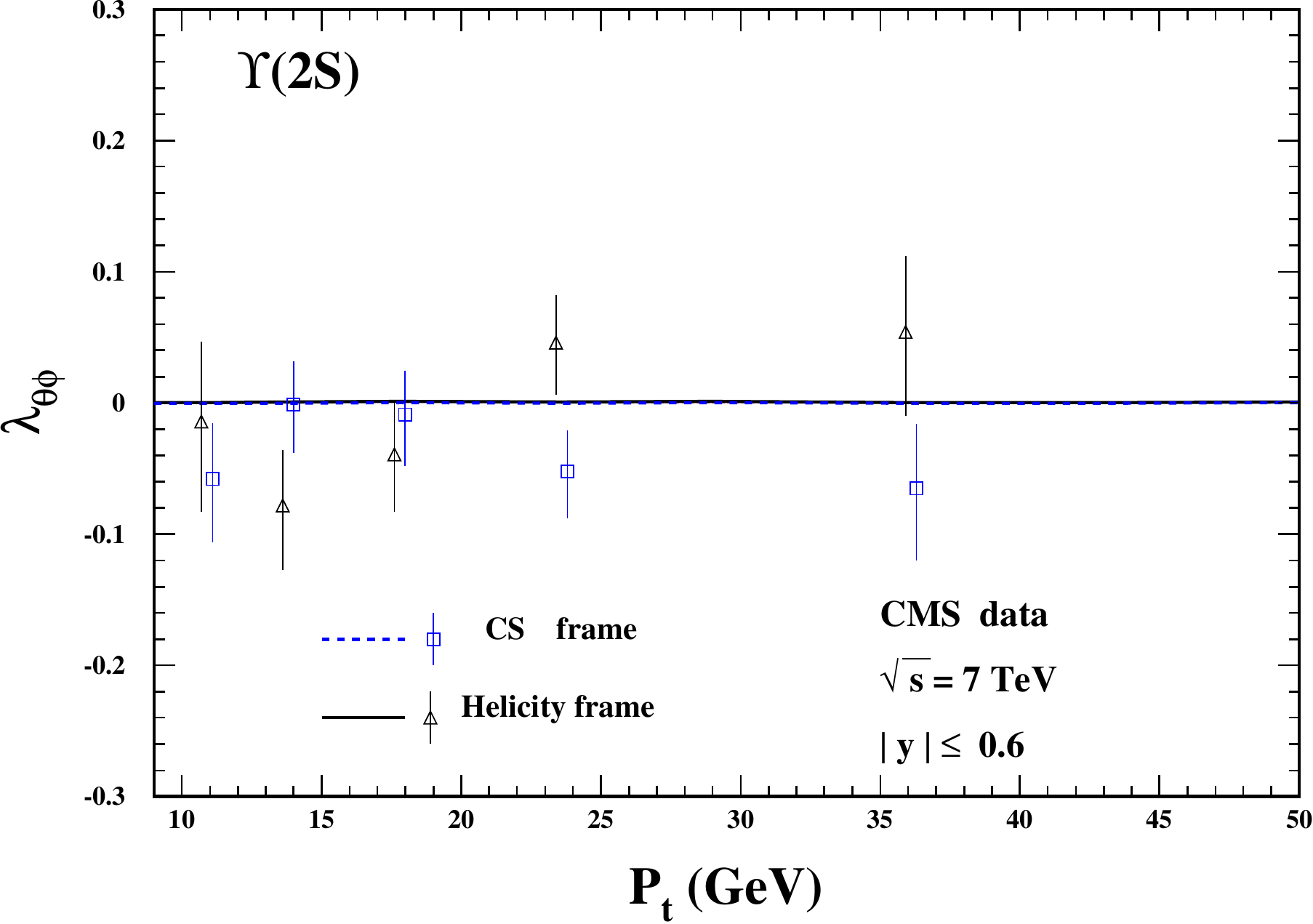}
  \includegraphics[width=5.0cm]{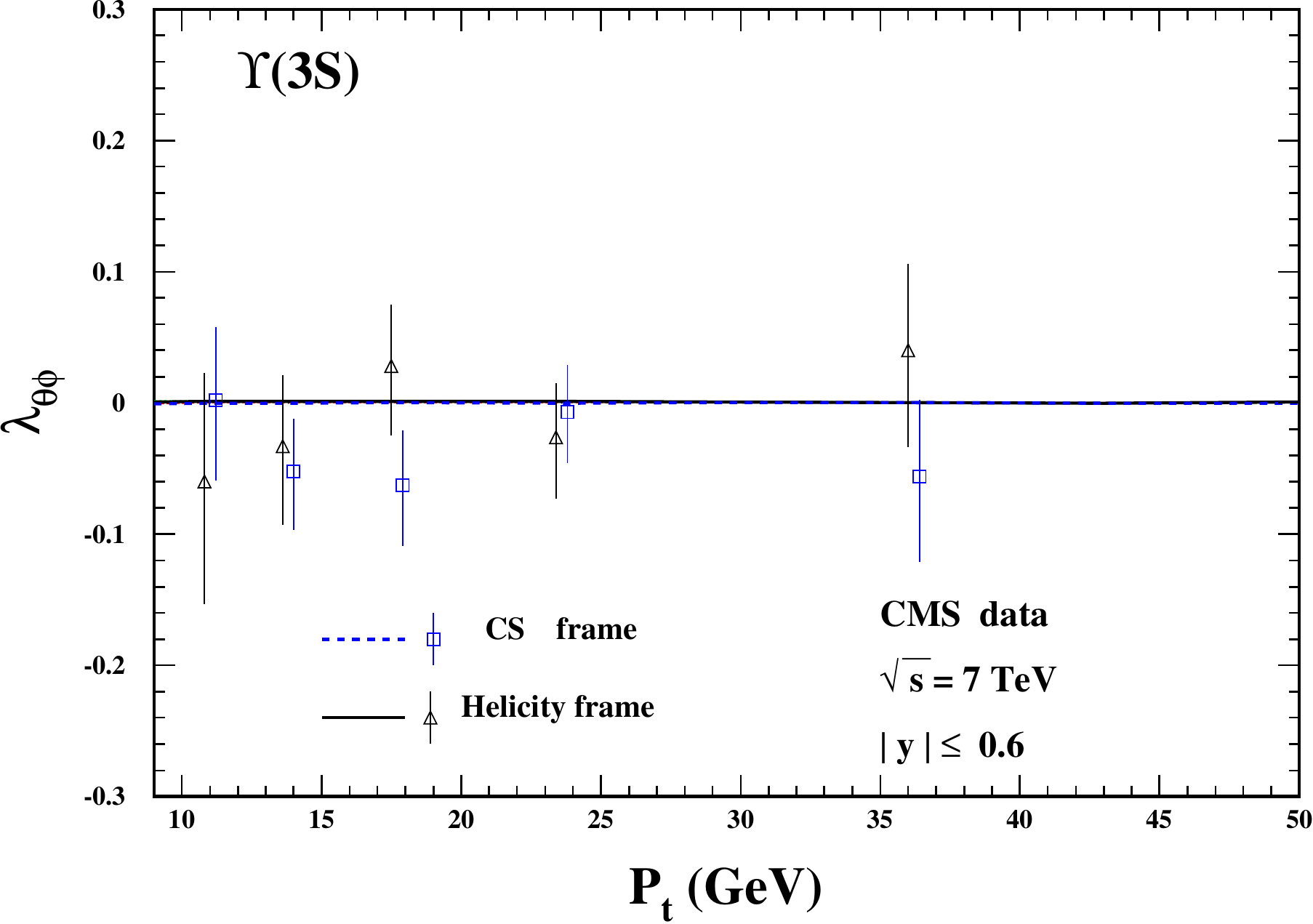} \\
  \includegraphics[width=5.0cm]{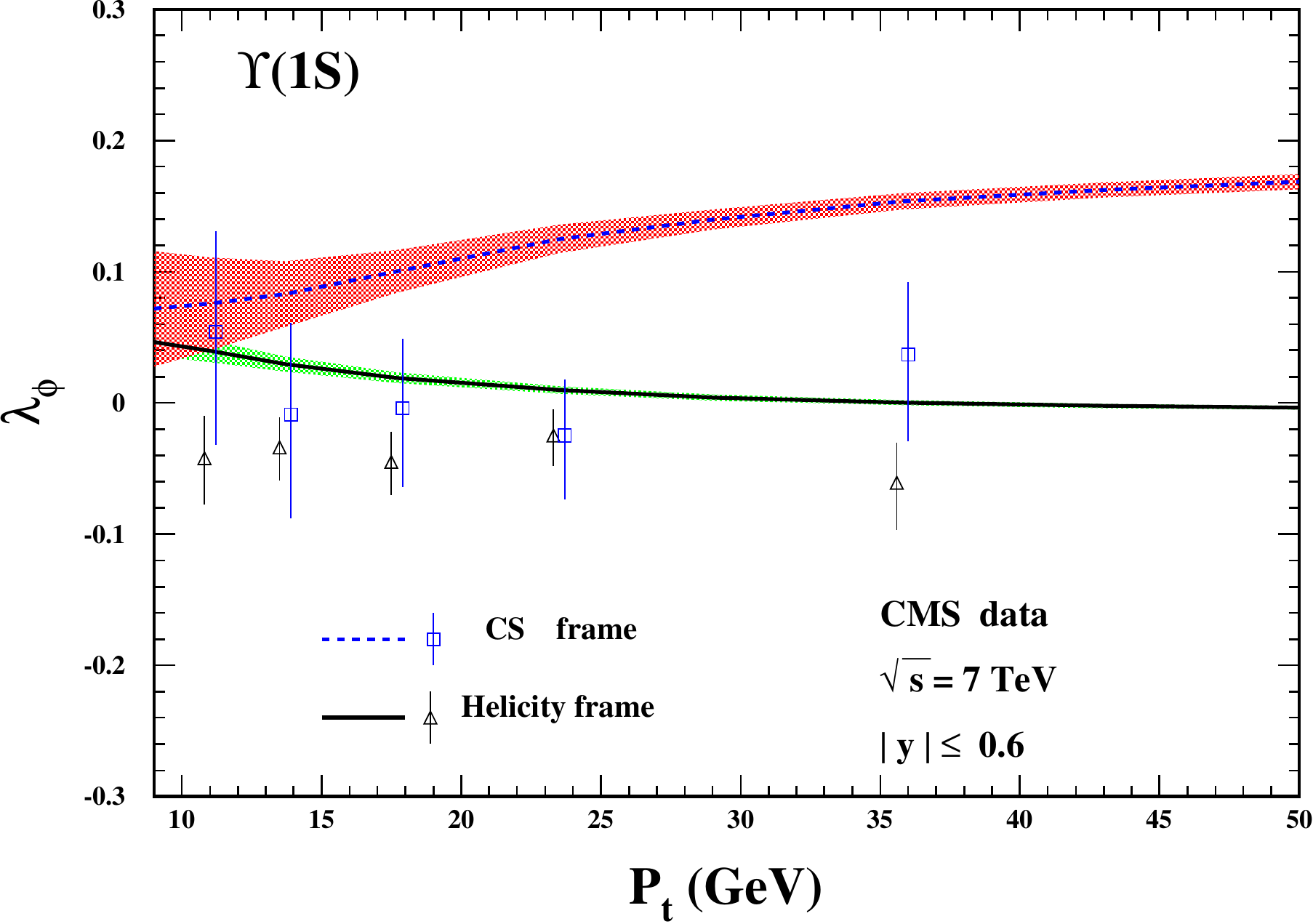}
  \includegraphics[width=5.0cm]{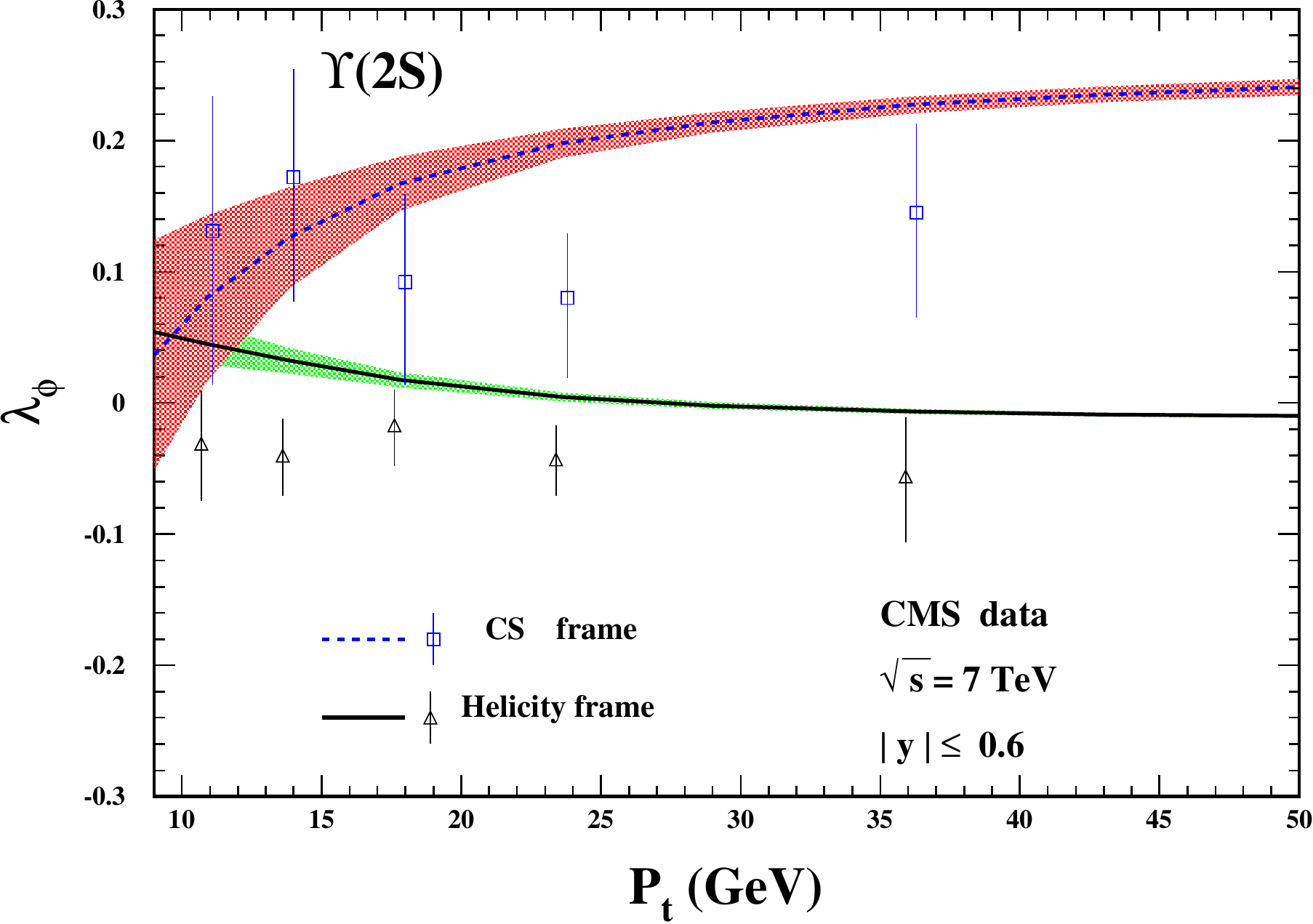}
  \includegraphics[width=5.0cm]{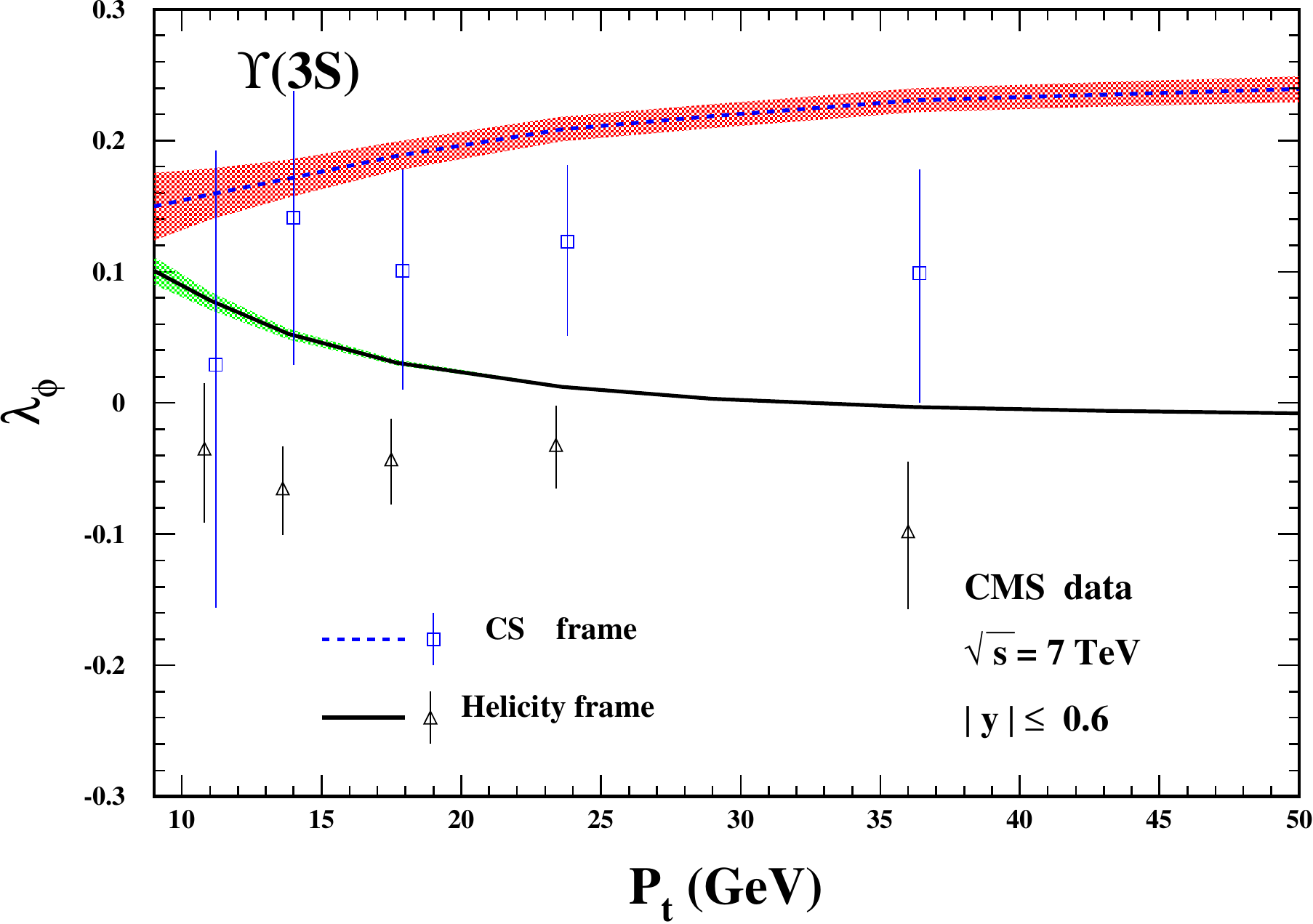} \\
  \caption{Polarization parameters $\lambda_{\theta}$(top), $\lambda_{\theta\phi}$(middle) and $\lambda_{\phi}$(bottom) for $\Upsilon$ hadroproduction in the rapidity region $|y| \leq 0.6$. From left to right:
  $\Upsilon(1S)$, $\Upsilon(2S)$ and $\Upsilon(3S)$. The CMS data is from Ref.~\cite{Chatrchyan:2012woa}.
   }
  \label{fig:pol:cms06}
\end{figure*}

\begin{figure*}[htb]
  \centering
  \includegraphics[width=5.0cm]{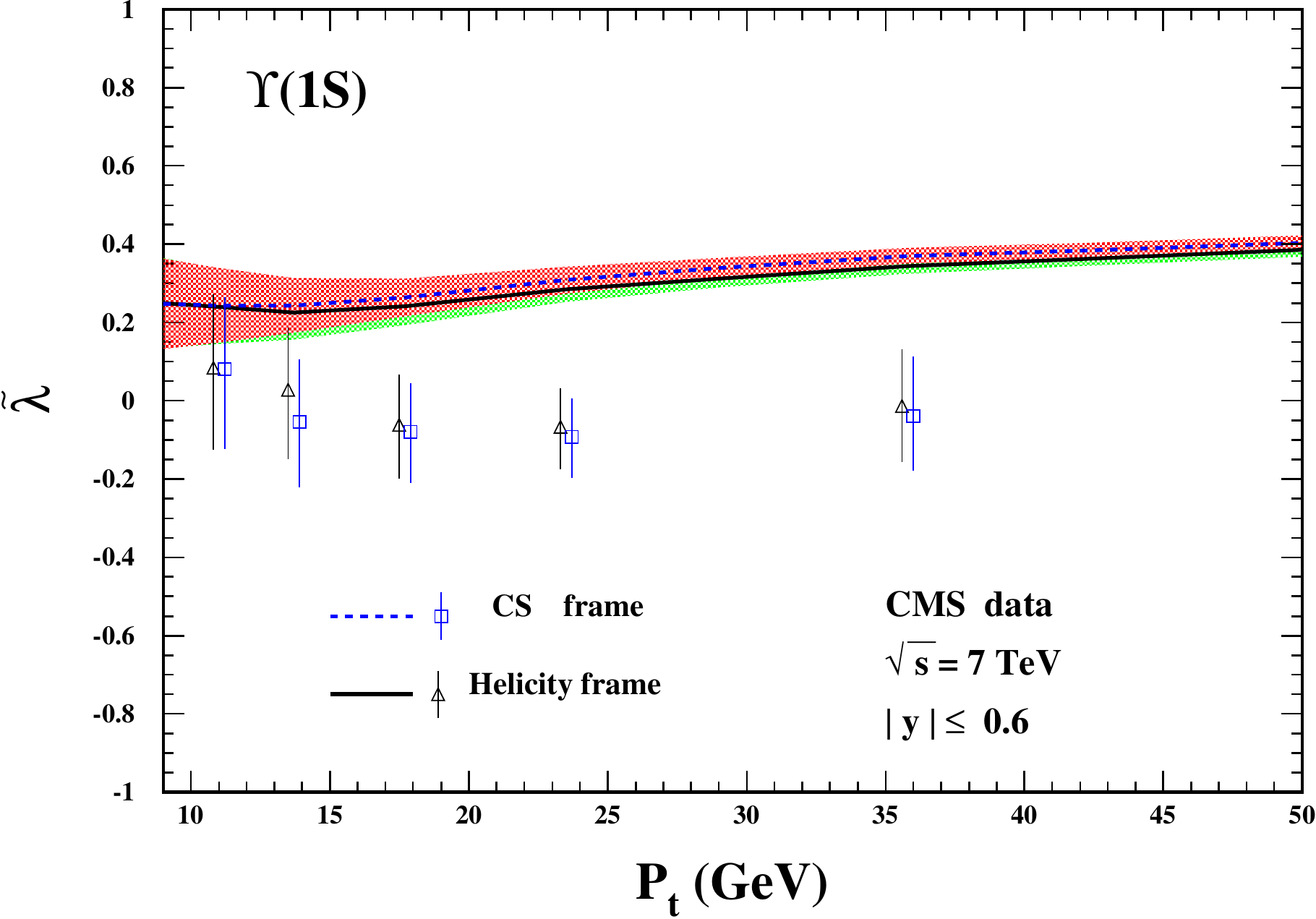}
  \includegraphics[width=5.0cm]{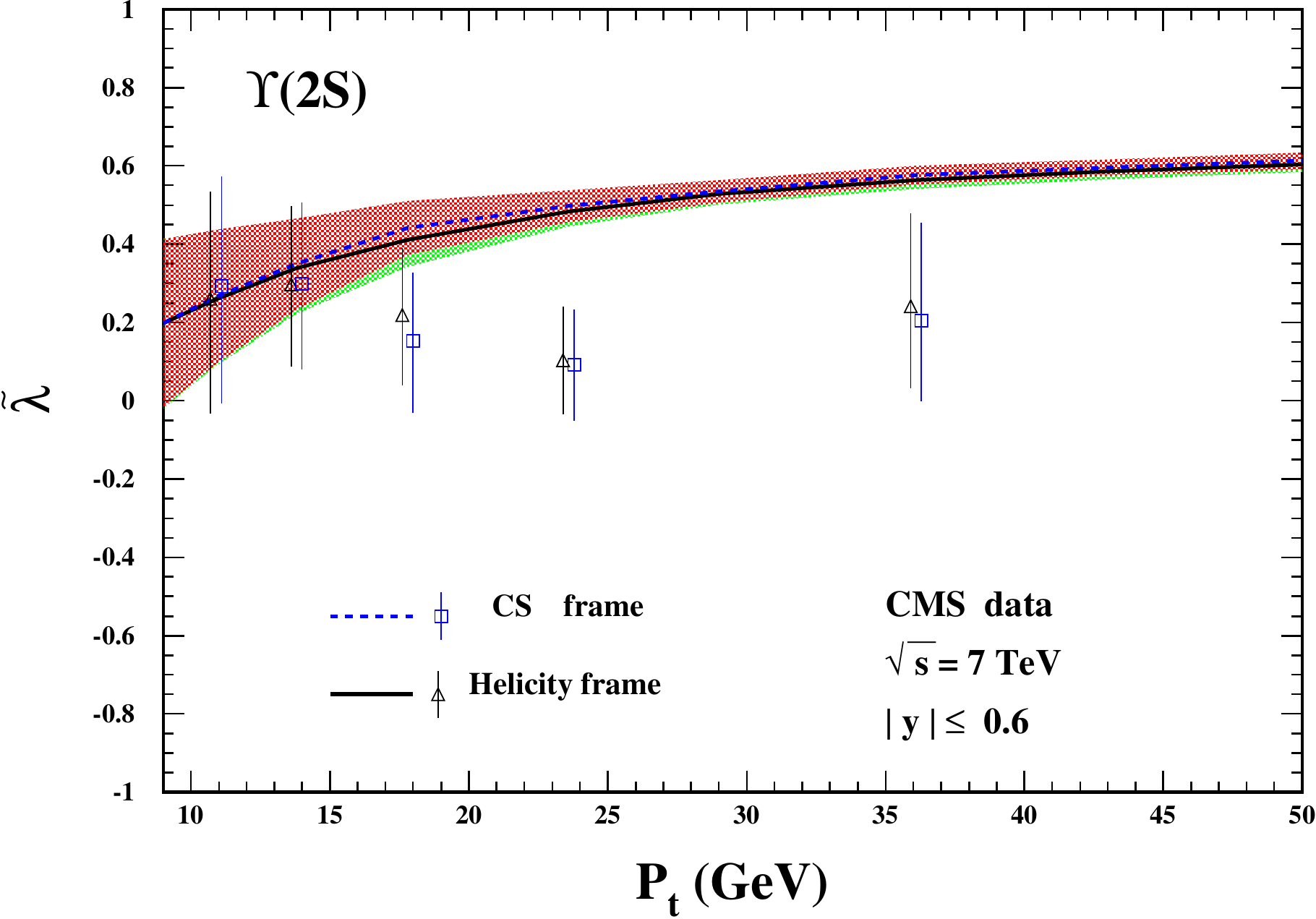}
  \includegraphics[width=5.0cm]{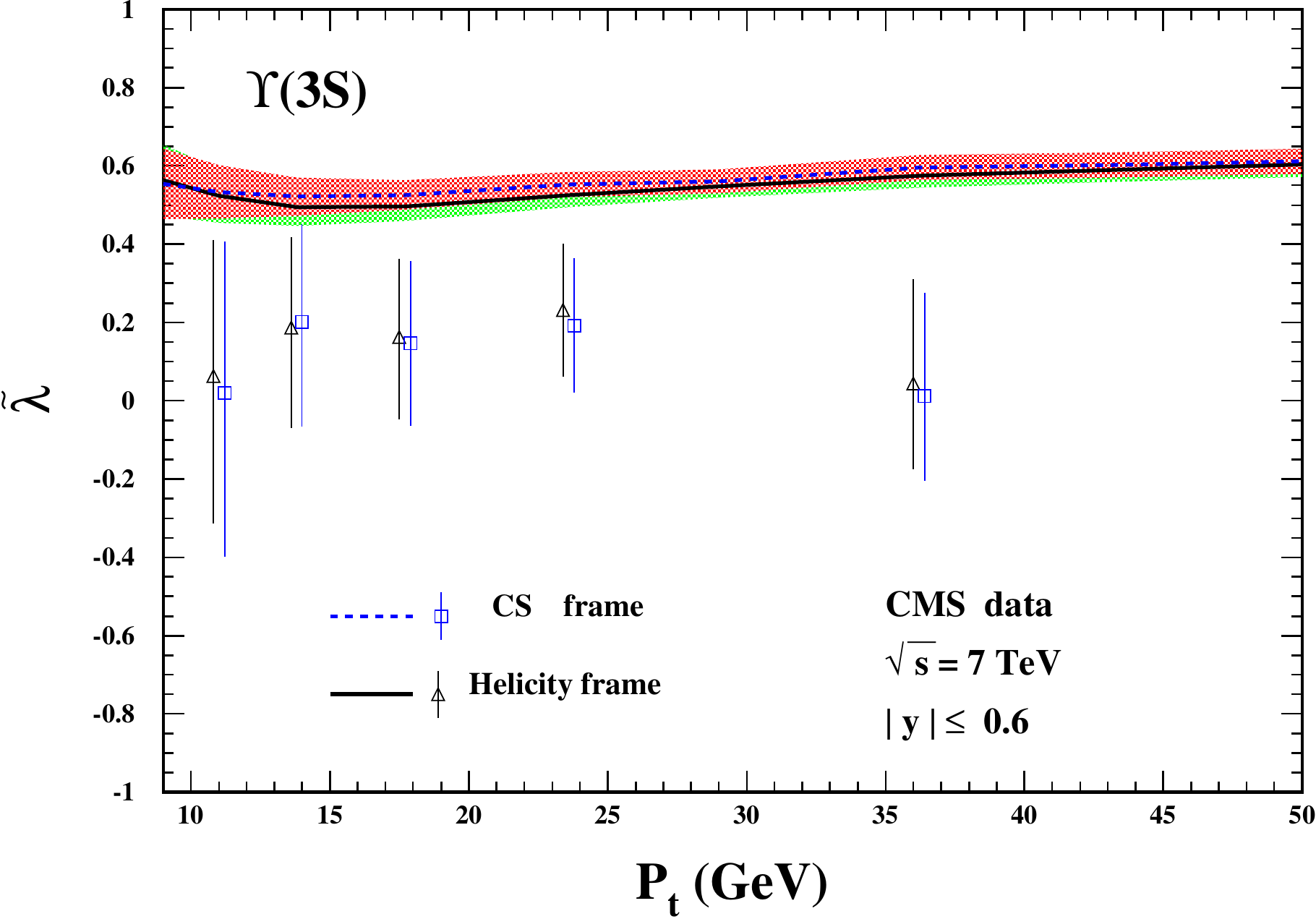}
  \caption{The frame-invariant quantity $\widetilde{\lambda}$ as functions of the $\Upsilon(1S,2S,3S)$'s transverse momentum $p_T$. The CMS data is from Ref.~\cite{Chatrchyan:2012woa}.
  }
  \label{fig:lamTilde:cms06}
\end{figure*}

\begin{figure*}[htb]
  \centering
  \includegraphics[width=5.0cm]{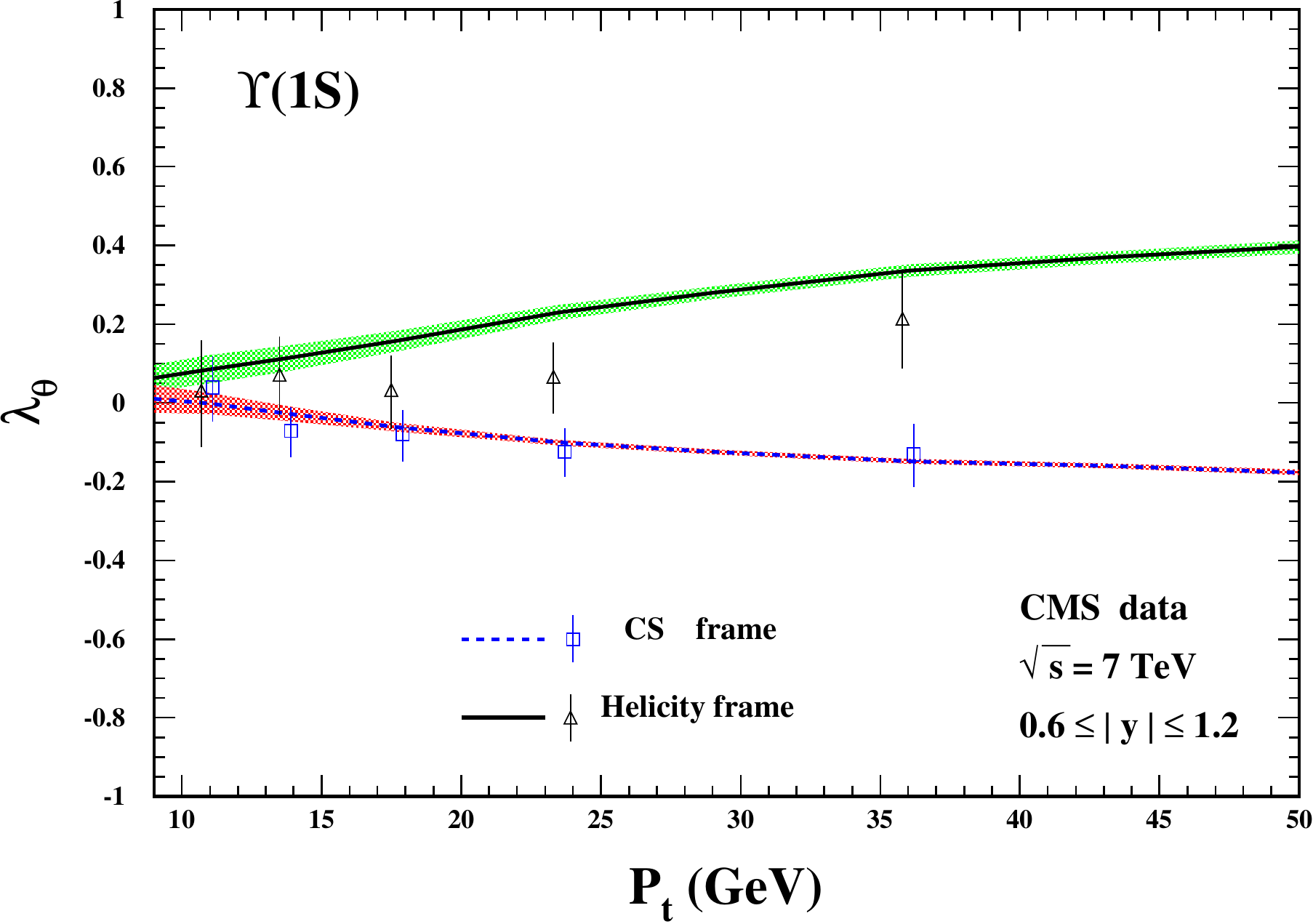}
  \includegraphics[width=5.0cm]{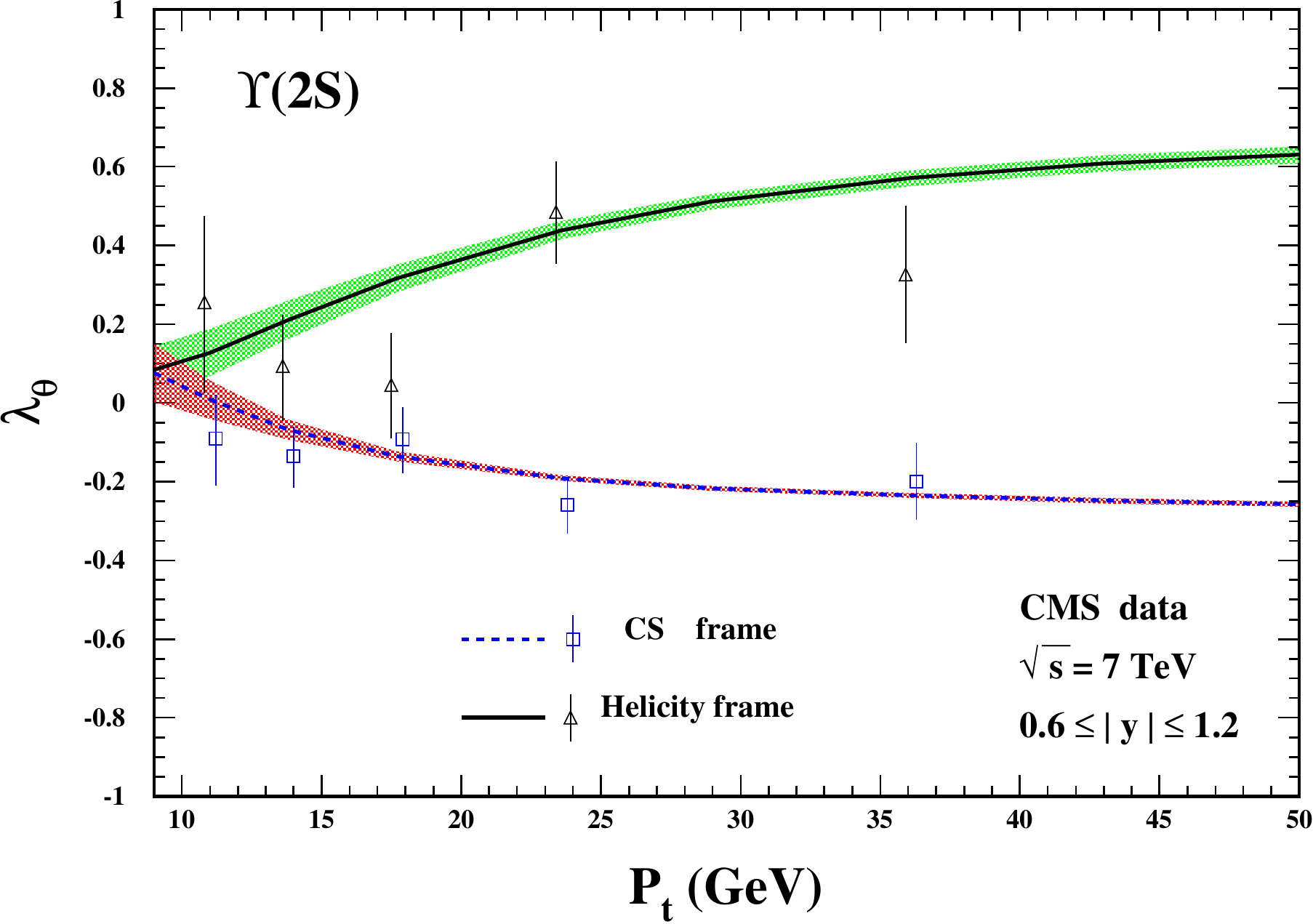}
  \includegraphics[width=5.0cm]{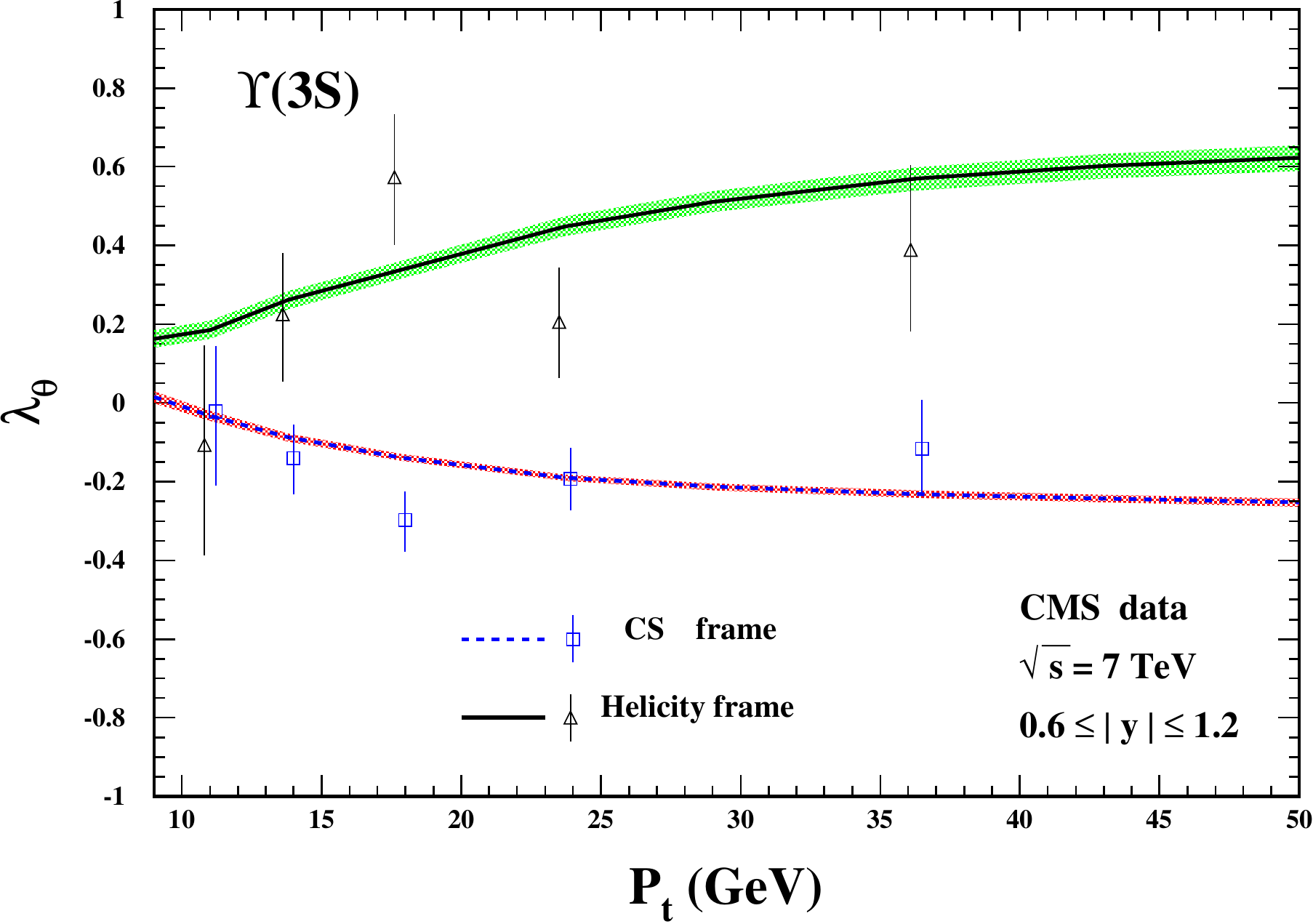} \\
  \includegraphics[width=5.0cm]{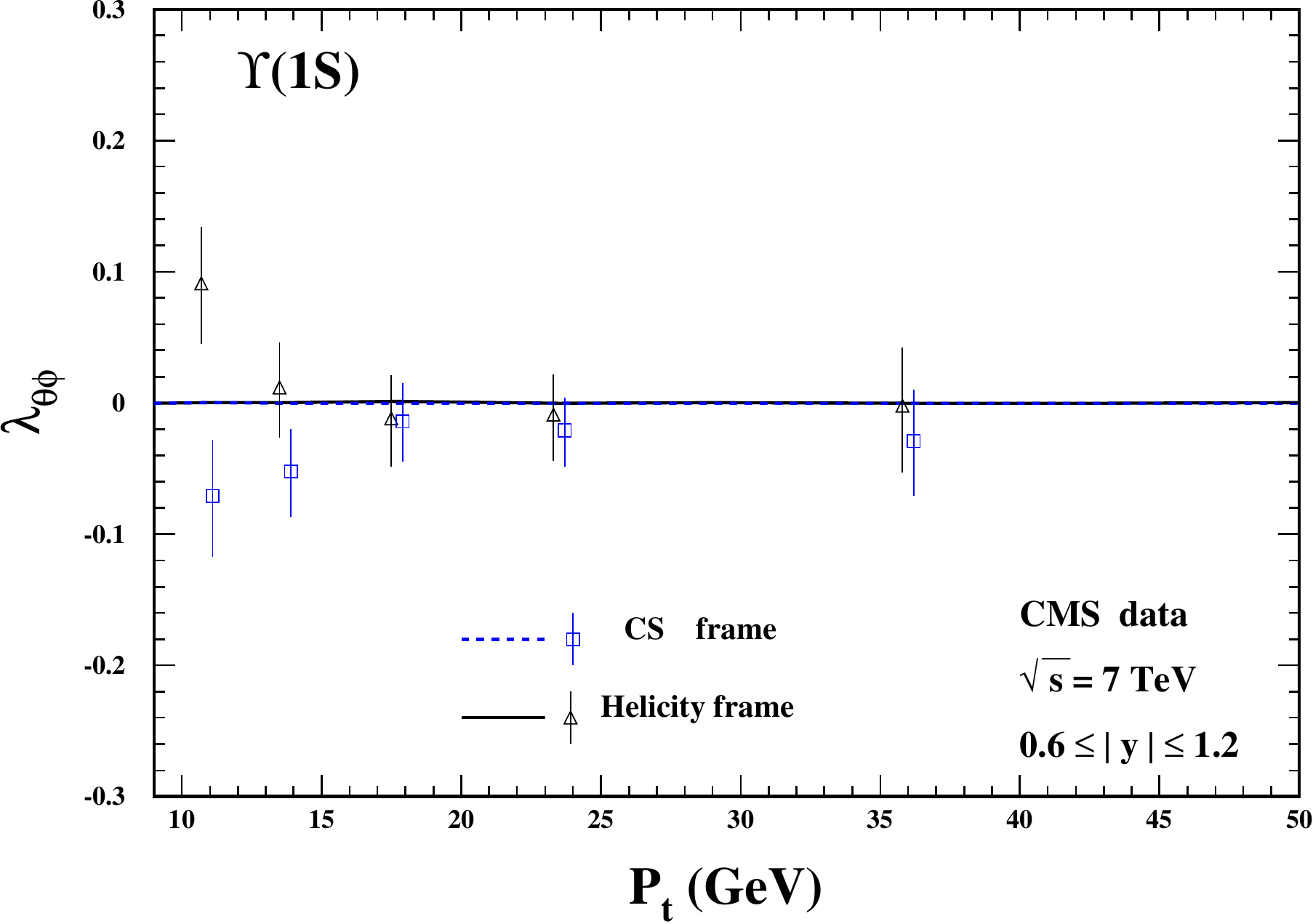}
  \includegraphics[width=5.0cm]{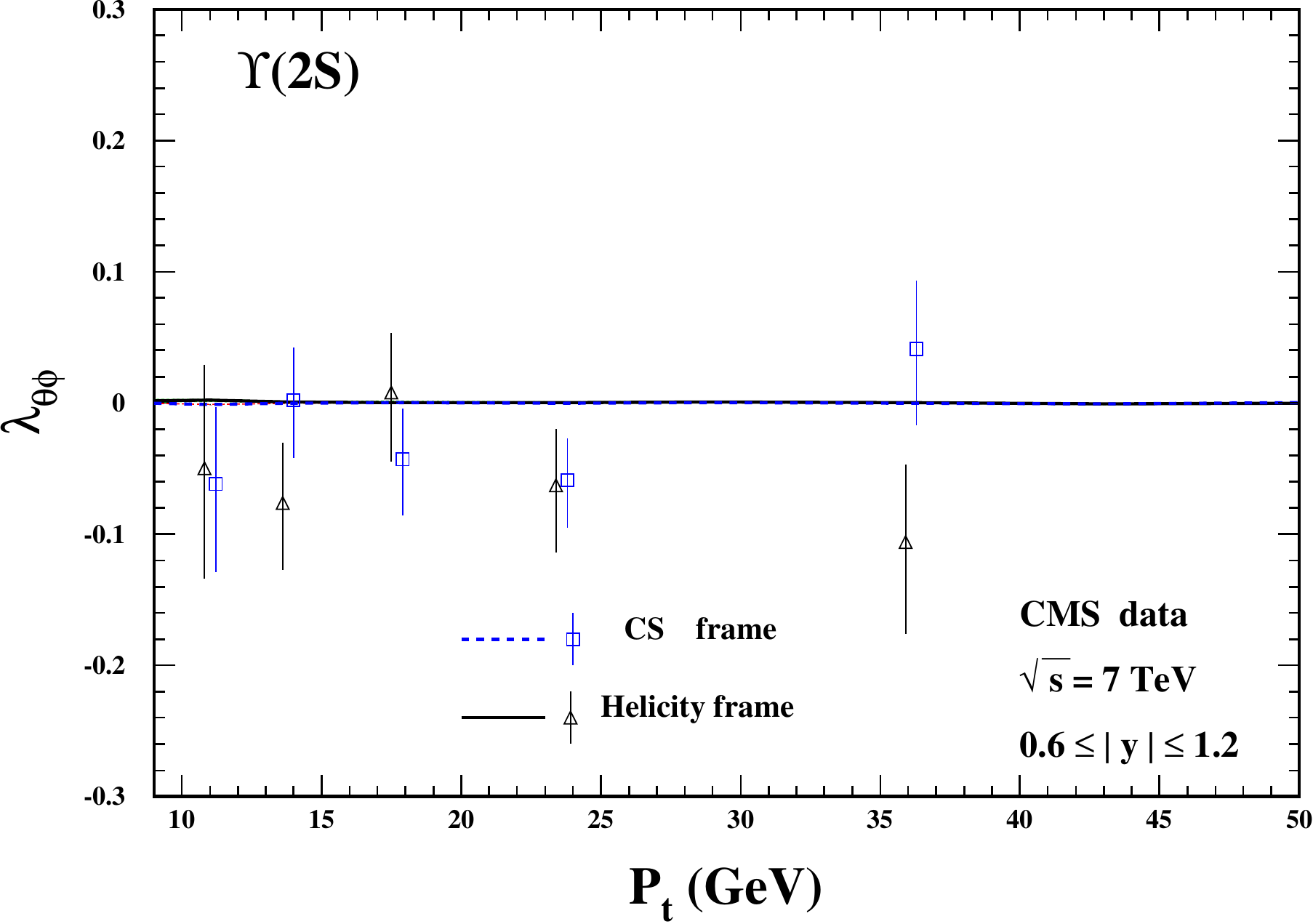}
  \includegraphics[width=5.0cm]{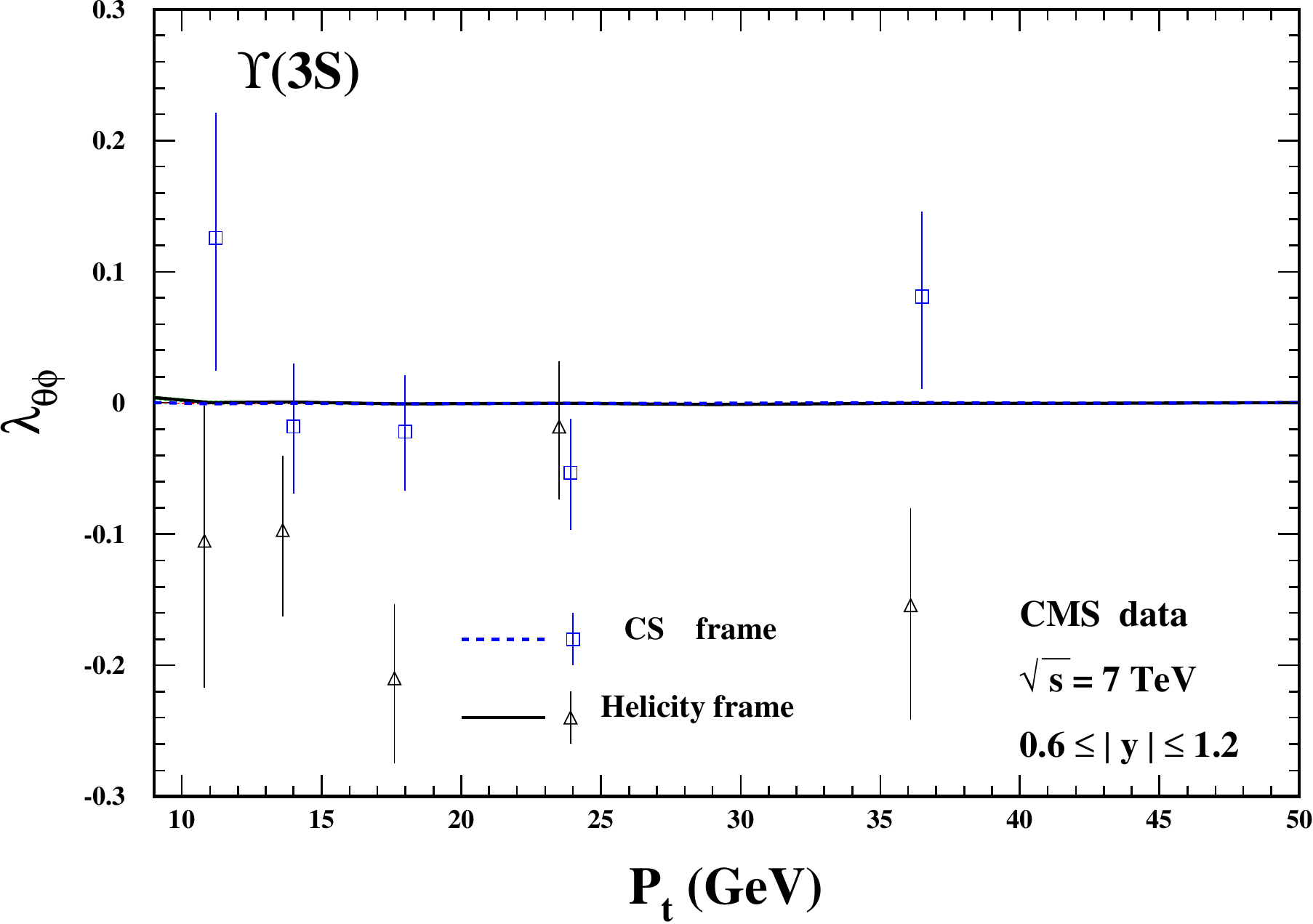} \\
  \includegraphics[width=5.0cm]{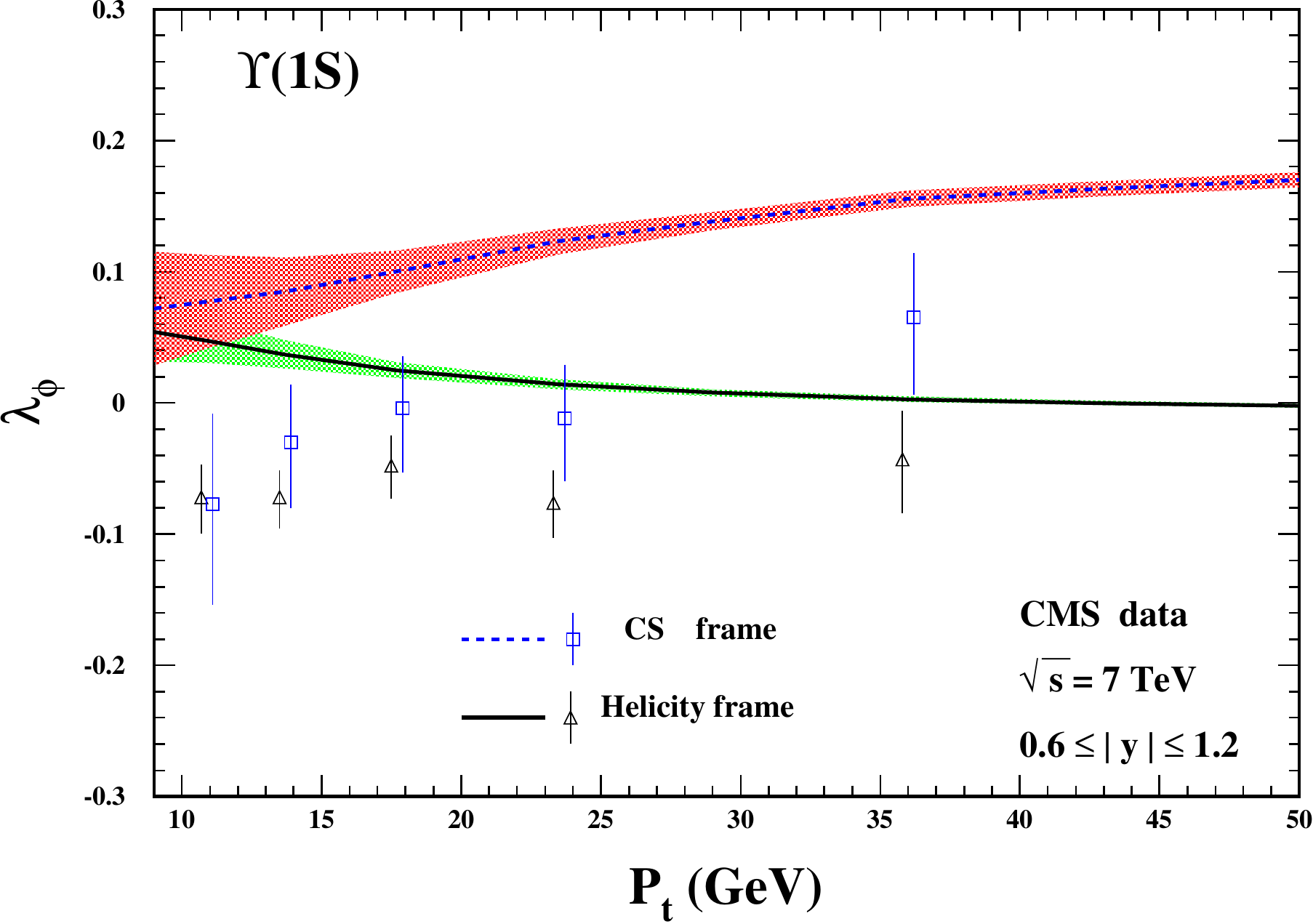}
  \includegraphics[width=5.0cm]{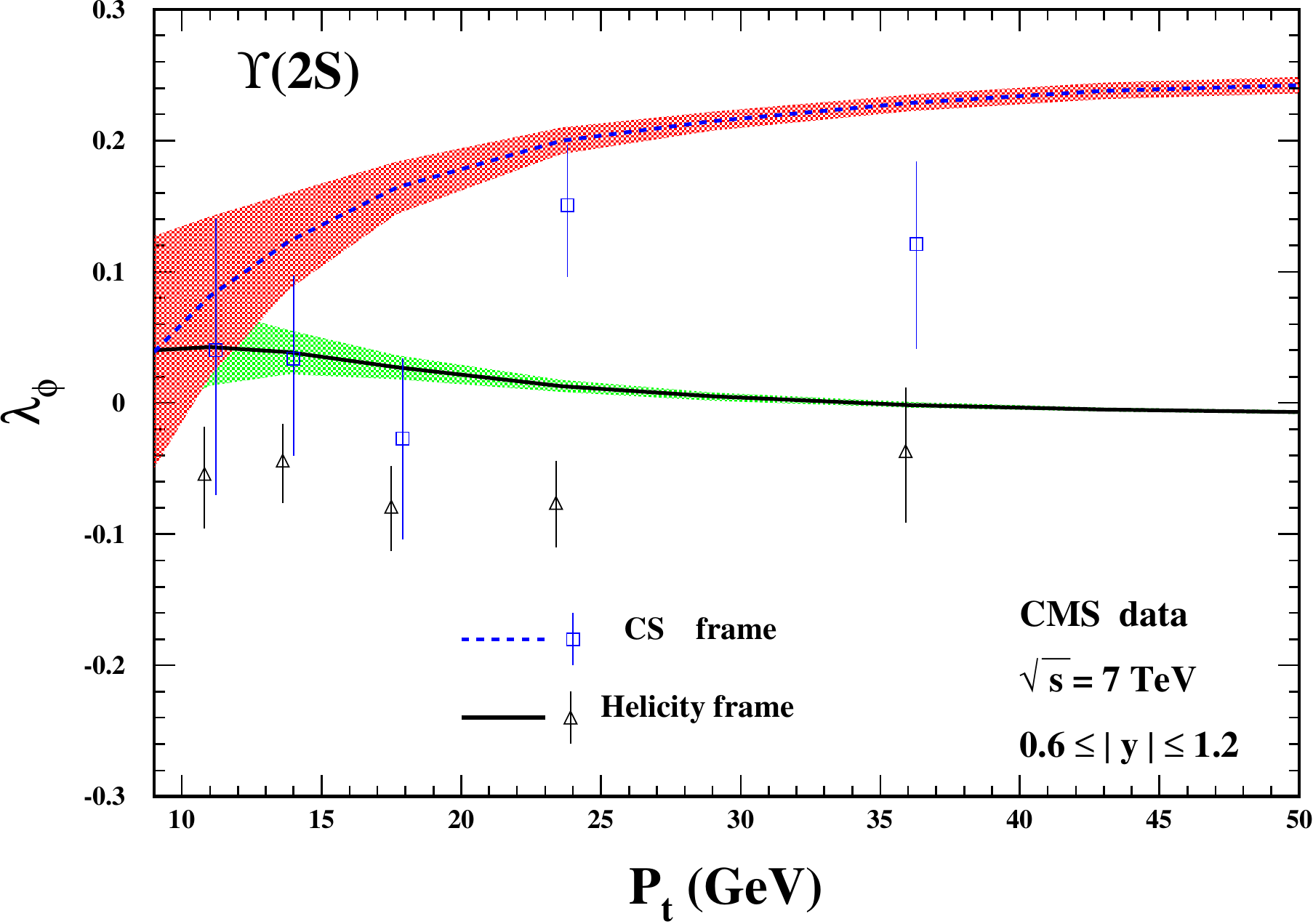}
  \includegraphics[width=5.0cm]{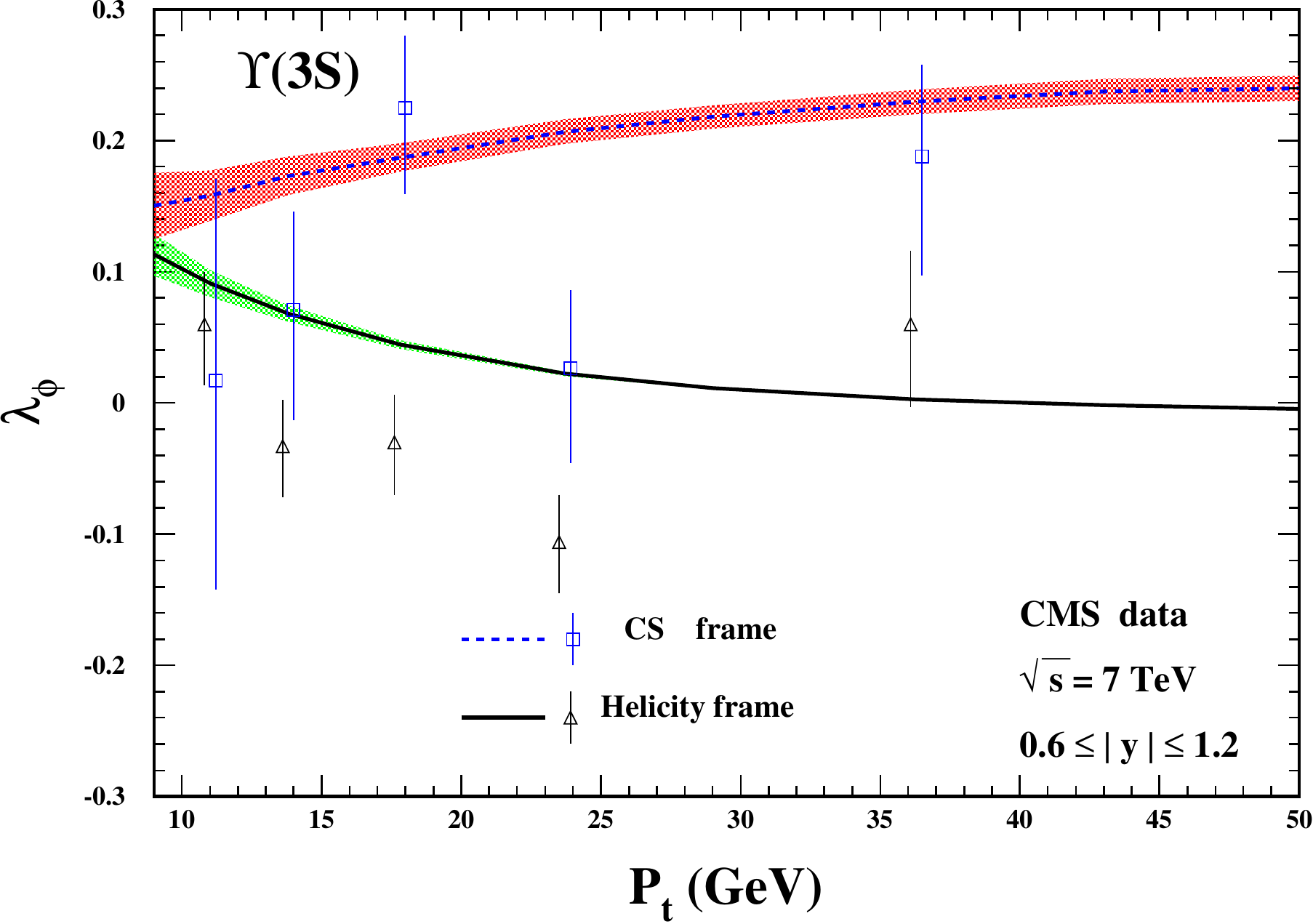} \\
  \caption{The same as Fig.~\ref{fig:pol:cms06} for the rapidity region $0.6 \leq |y| \leq 1.2 $ at CMS~\cite{Chatrchyan:2012woa}.}
  \label{fig:pol:cms12}
\end{figure*}

\begin{figure*}[htb]
  \centering
  \includegraphics[width=5.0cm]{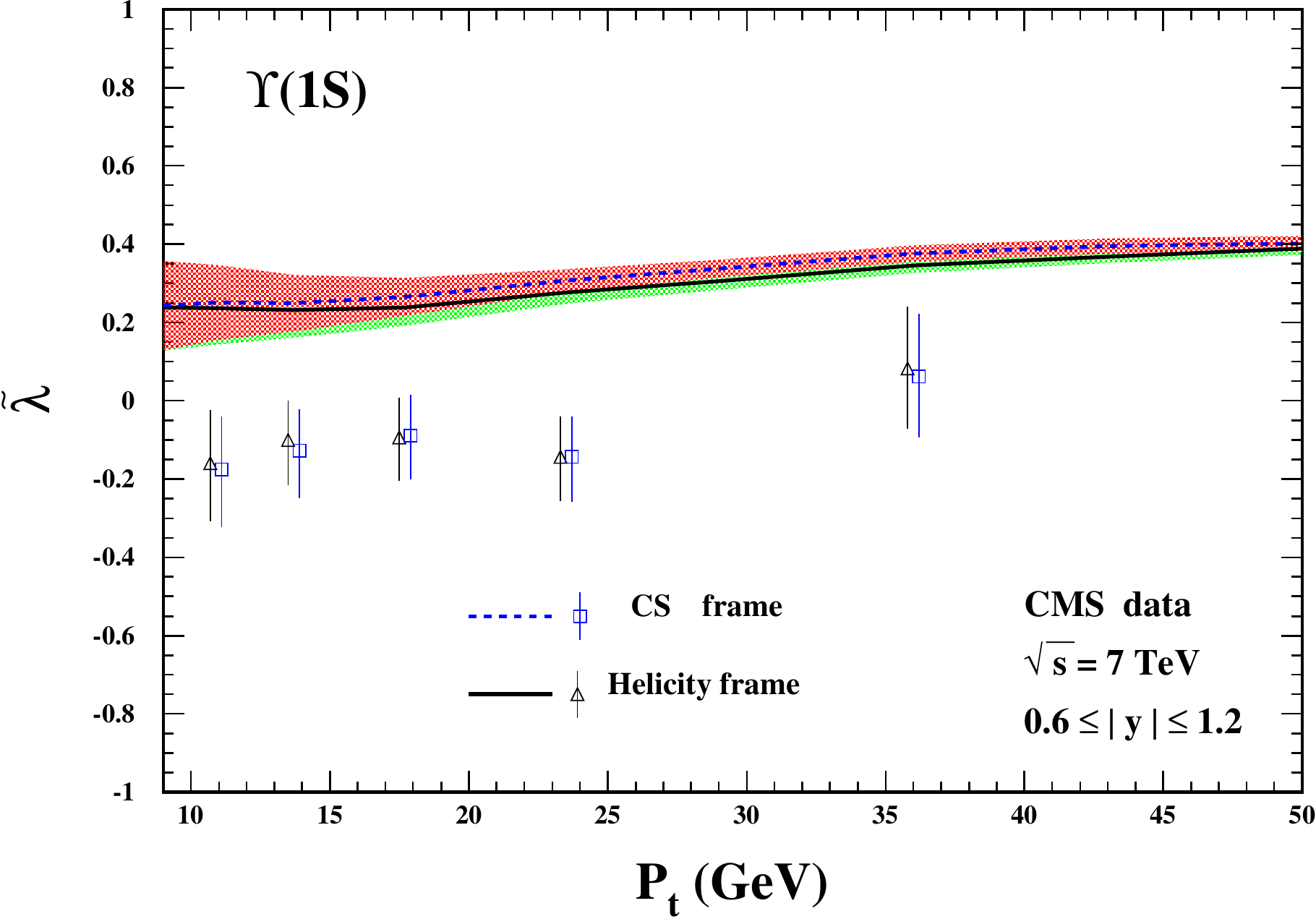}
  \includegraphics[width=5.0cm]{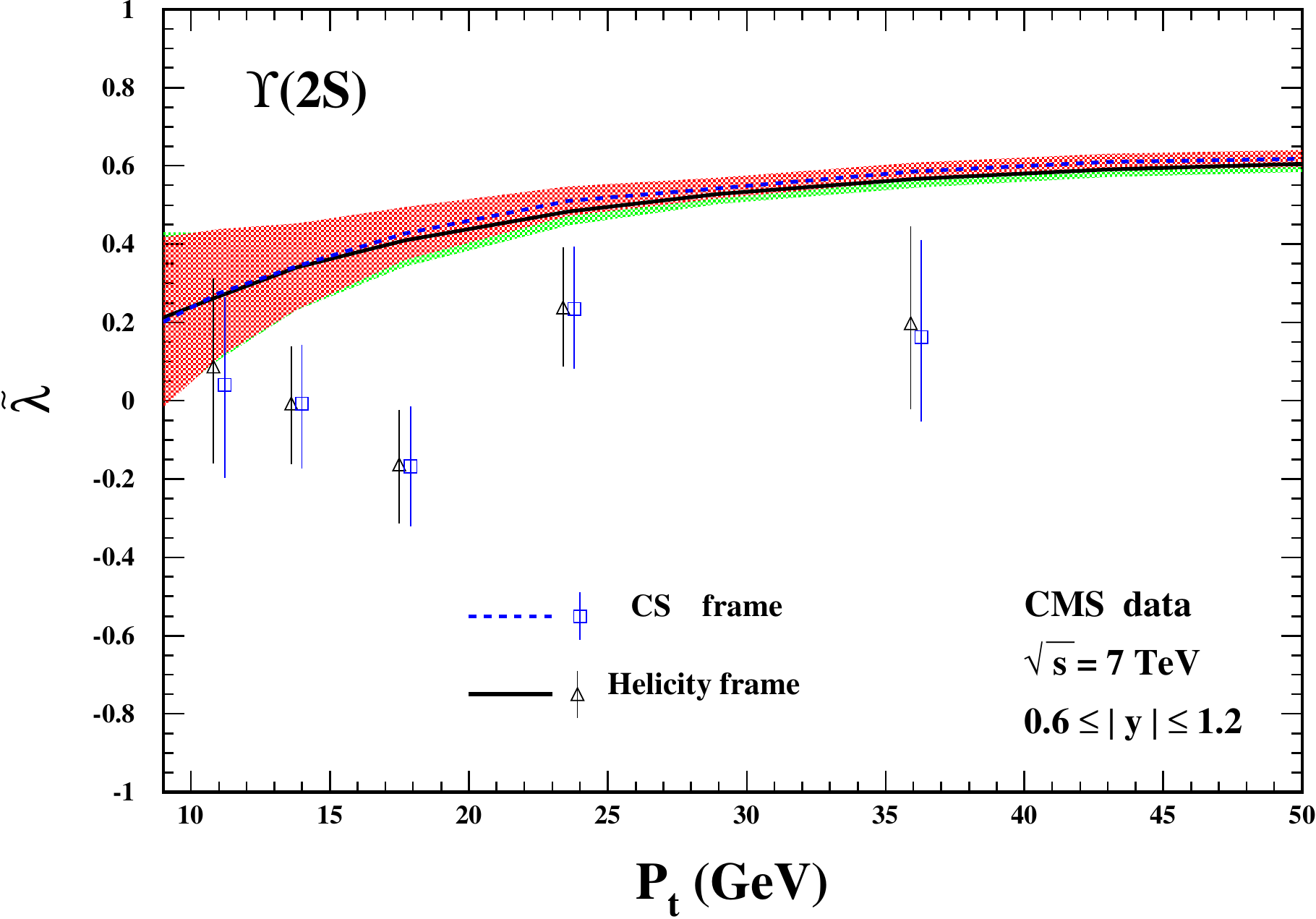}
  \includegraphics[width=5.0cm]{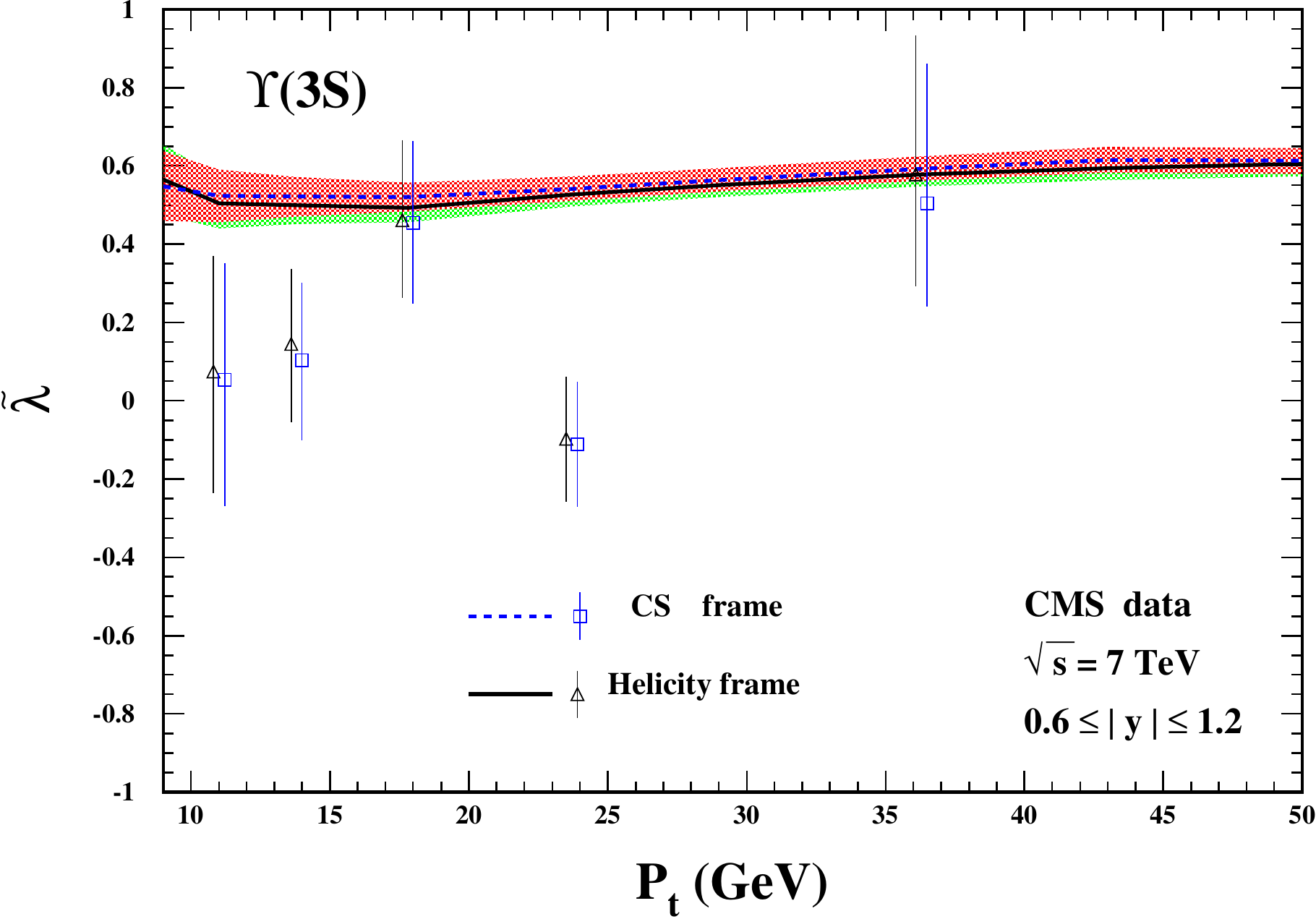}
  \caption{The frame-invariant quantity $\widetilde{\lambda}$ for $\Upsilon(1S,2S,3S)$ hadroproduction
  in the rapidity region $0.6 \leq |y| \leq 1.2 $. The CMS data is from Ref.~\cite{Chatrchyan:2012woa}.
   }
  \label{fig:lamTilde:cms12}
\end{figure*}

The predictions of $\Upsilon(1S,2S,3S)$ polarization parameters in the rapidity region $|y| \leq 0.6$ relating to CMS measurements
are computed and presented in Figs.~\ref{fig:pol:cms06} and \ref{fig:lamTilde:cms06},
for ($\lambda_{\theta}$, $\lambda_{\theta\phi}$, $\lambda_{\phi}$) and $\widetilde{\lambda}$, respectively.
In Fig.~\ref{fig:pol:cms06}, $\lambda_{\theta}$ in the helicity frame is renewed,
which is found to be the same as shown in Fig. 2 of Ref.~\cite{Feng:2015wka} where the corresponding data was used to extract the LDMEs in Table~\ref{tab:ldme}.
The prediction of $\lambda_{\theta}$ in the CS frame,
which denoted by the blue-dotted lines, are consistent with all the data for $\Upsilon(1S,2S,3S)$.
$\lambda_{\theta\phi}$, which was investigated for $J/\psi$ in Ref.~\cite{Feng:2018ukp},
is exactly zero in the symmetric rapidity region in the helicity frame.
In addition, here $\lambda_{\theta\phi}$ is also zero in the CS frame.

$\lambda_{\phi}$ of $\Upsilon(nS)$ behaves in a similar way in both the helicity and the CS frames.
The theoretical results in the helicity frame almost describe all the data for $\Upsilon(2S,3S)$, while for $\Upsilon(1S)$ the theory and the experimental data deviated with different tendency.
This situation is much better for the prediction in the CS frame although there are still
small deviations between the theoretical curves and the corresponding experimental data.

In Fig.~\ref{fig:lamTilde:cms06}, we present the results for the frame invariant quantity $\widetilde{\lambda}$ defined in Eq.~(\ref{eq:laminv}).
It clearly shows that
our theoretical results in the helicity and CS frames are coincide with each other.
But there are small differences between the theoretical results and the corresponding experimental data.
The prediction of $\widetilde{\lambda}$ can cover about two data points in the lower $p_t$ region ($p_t < 15$ GeV) for all the three states $\Upsilon(1S,2S,3S)$.
While in the higher $p_t$ region, the theoretical results are higher than the experimental data but still within 2$\sigma$ accuracy.
Especially, one may notice that the theoretical predictions and the experimental data for $\Upsilon(2S)$ behave in a contrary way
with the increasing of the transverse momentum $p_t$.

As regards the $\Upsilon(nS)$ polarization in $0.6 \leq |y| \leq 1.2$ rapidity region,
they are similar with the cases in $|y| \leq 0.6$.
We omit the detail descriptions here and present the plots in Figs.~\ref{fig:pol:cms12} and \ref{fig:lamTilde:cms12}.

\subsection{The polarization relating to LHCb measurements}
\begin{figure*}[htb]
  \centering
  \includegraphics[width=5.0cm]{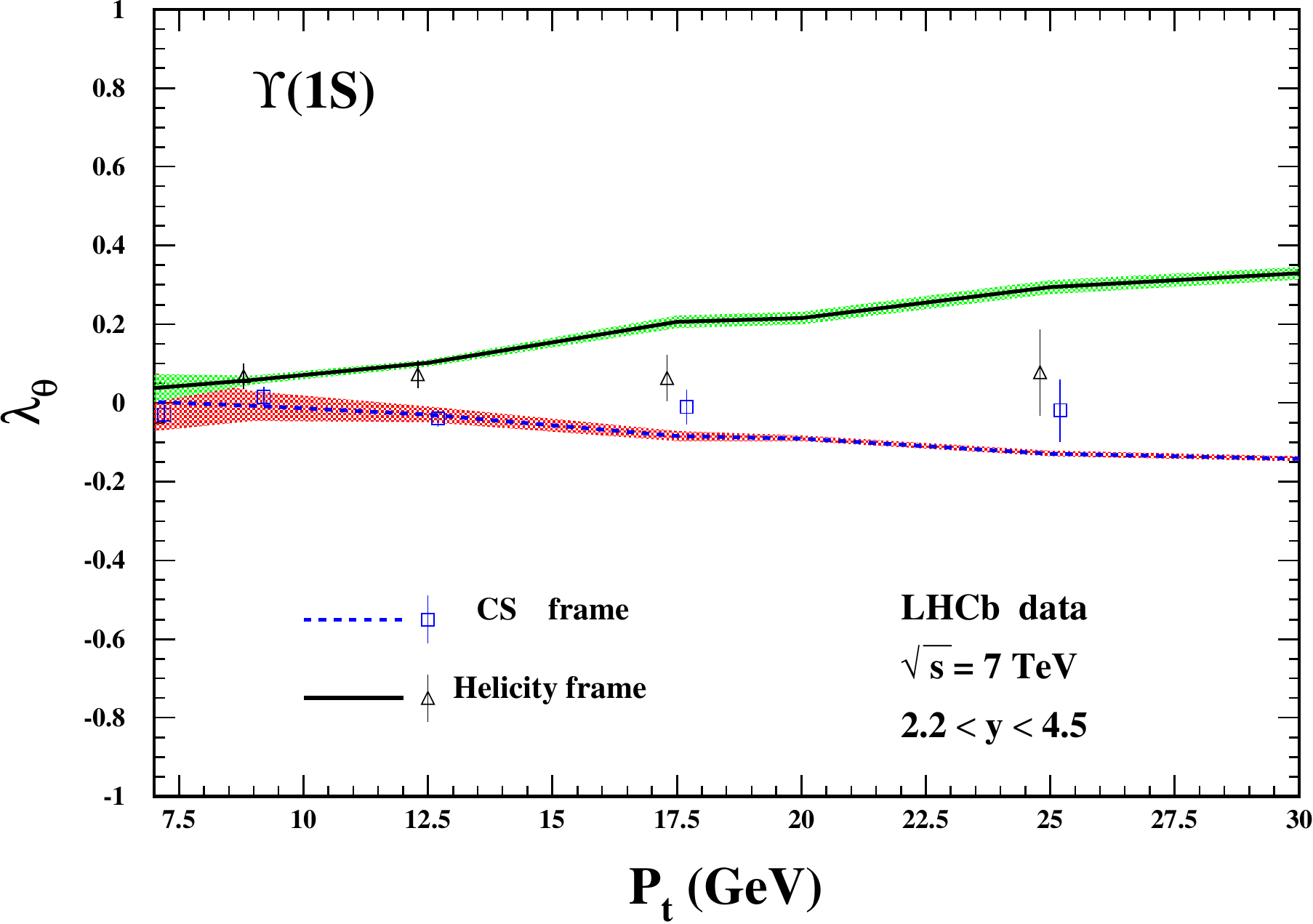}
  \includegraphics[width=5.0cm]{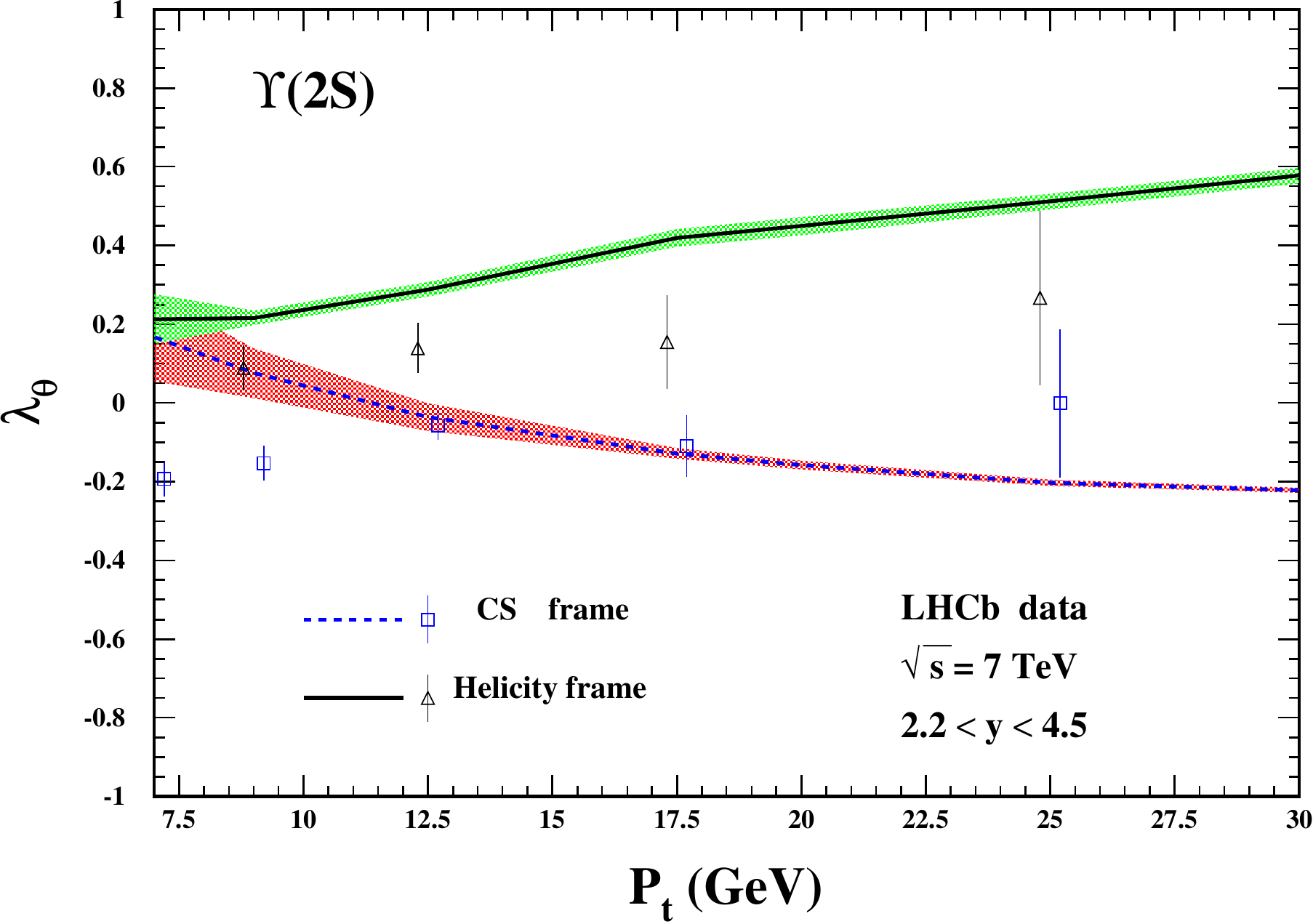}
  \includegraphics[width=5.0cm]{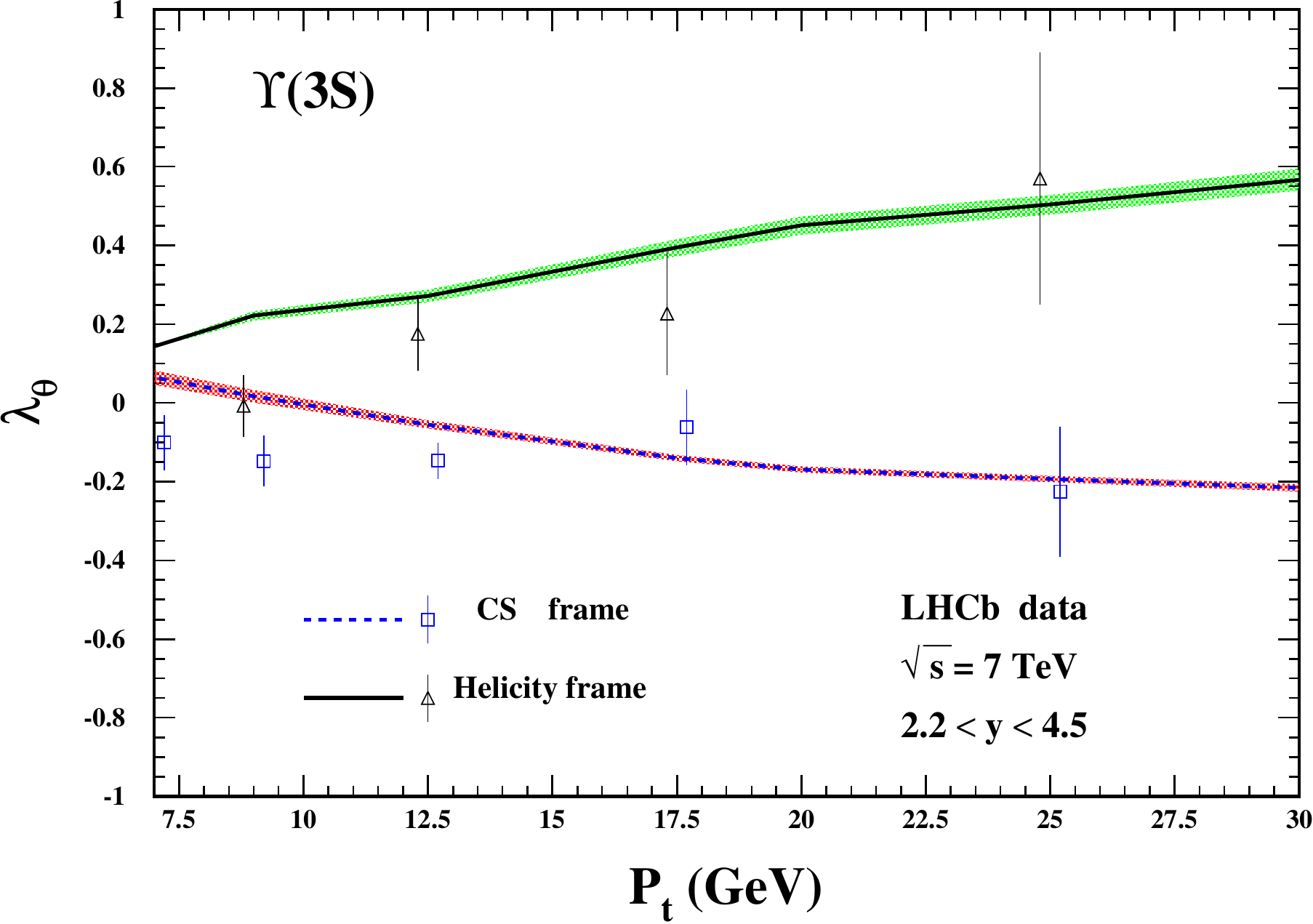} \\
  \includegraphics[width=5.0cm]{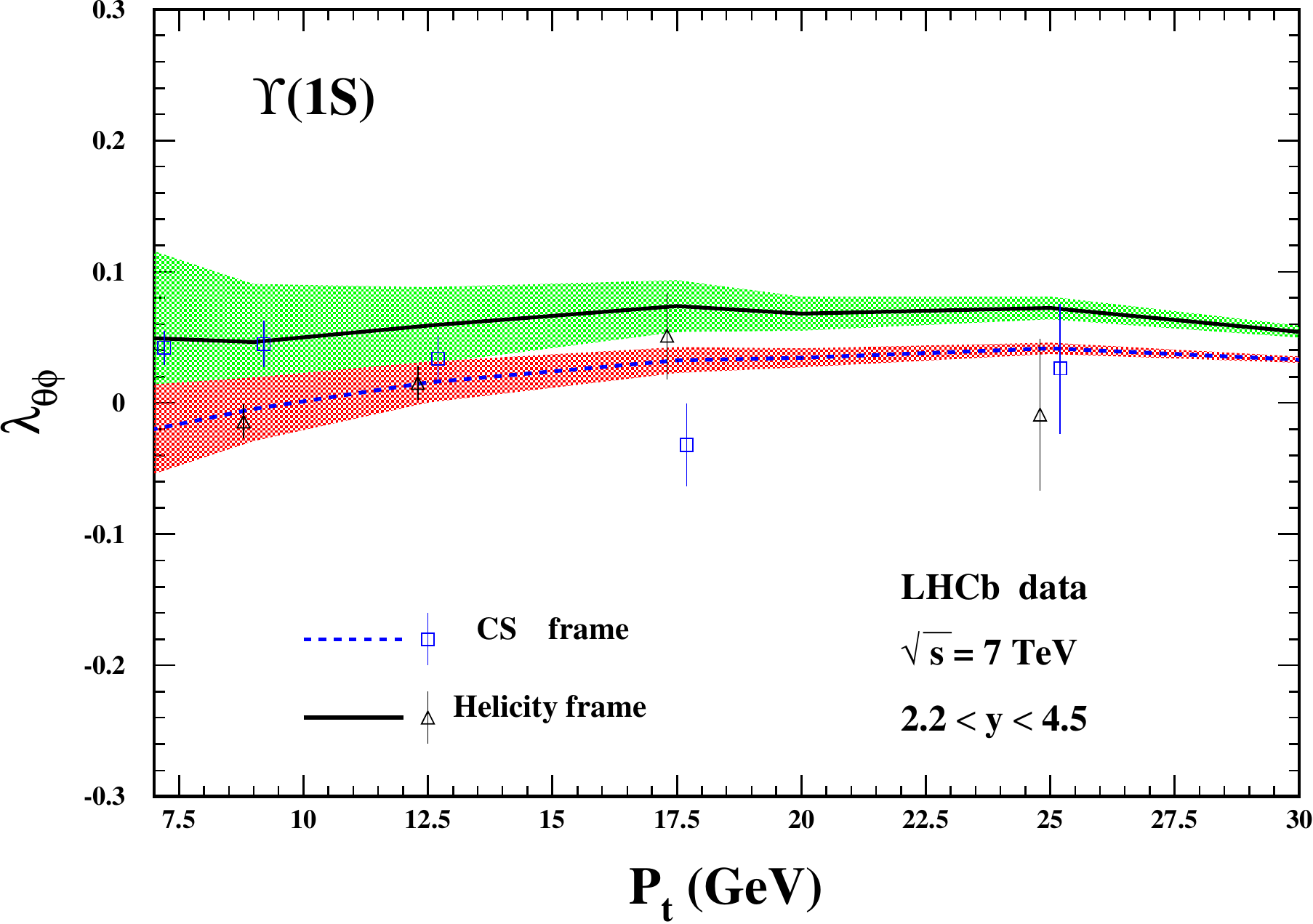}
  \includegraphics[width=5.0cm]{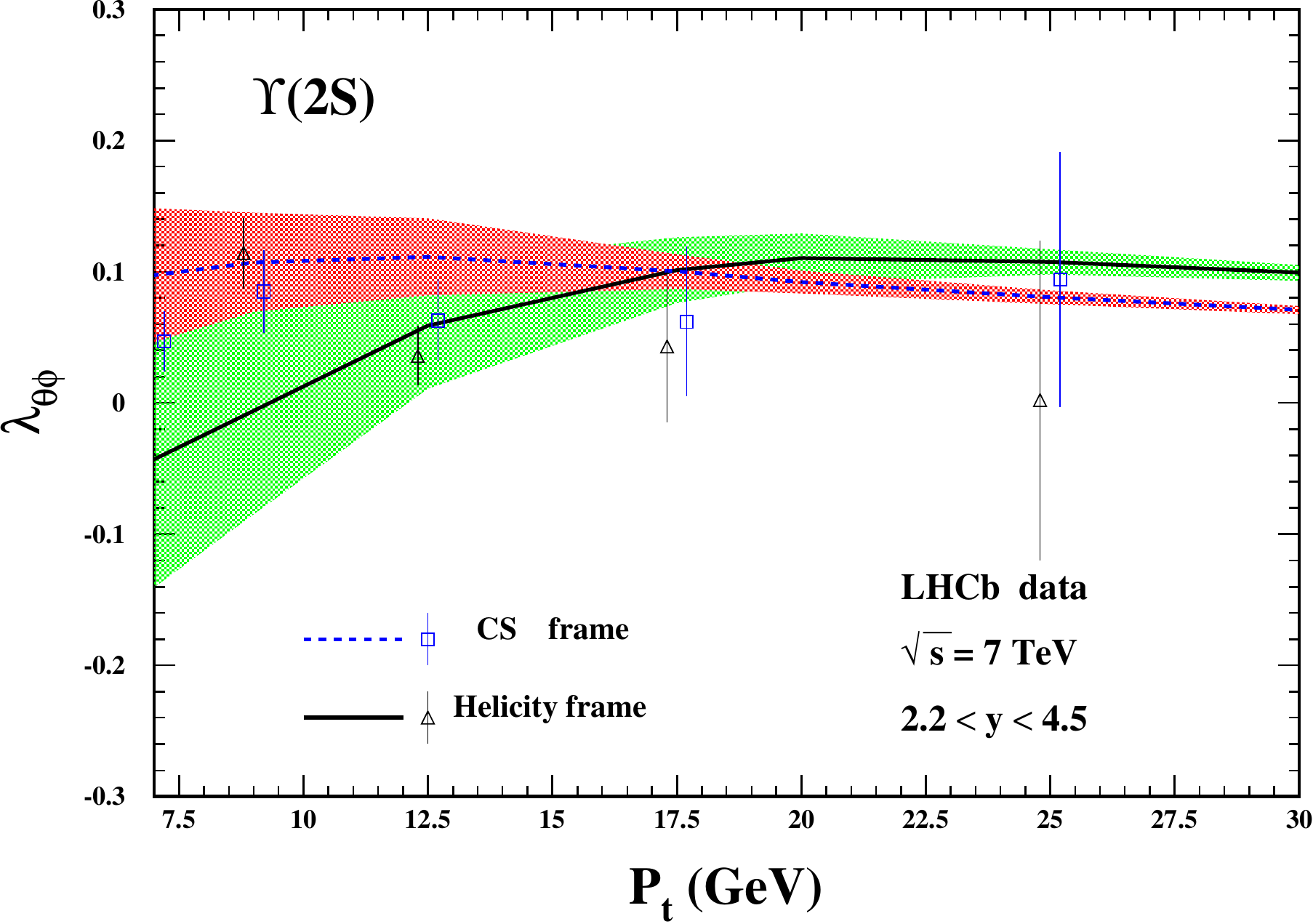}
  \includegraphics[width=5.0cm]{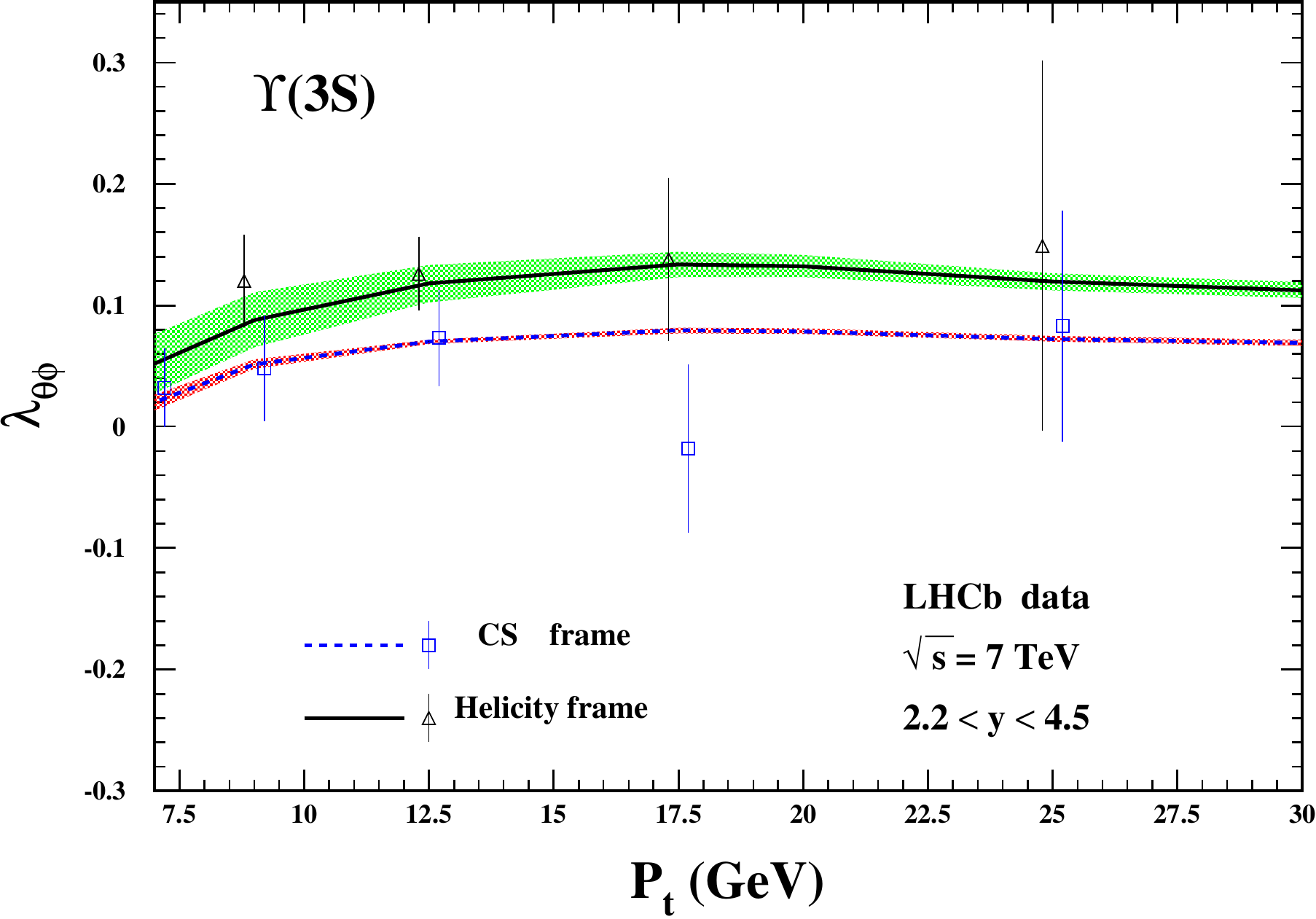} \\
  \includegraphics[width=5.0cm]{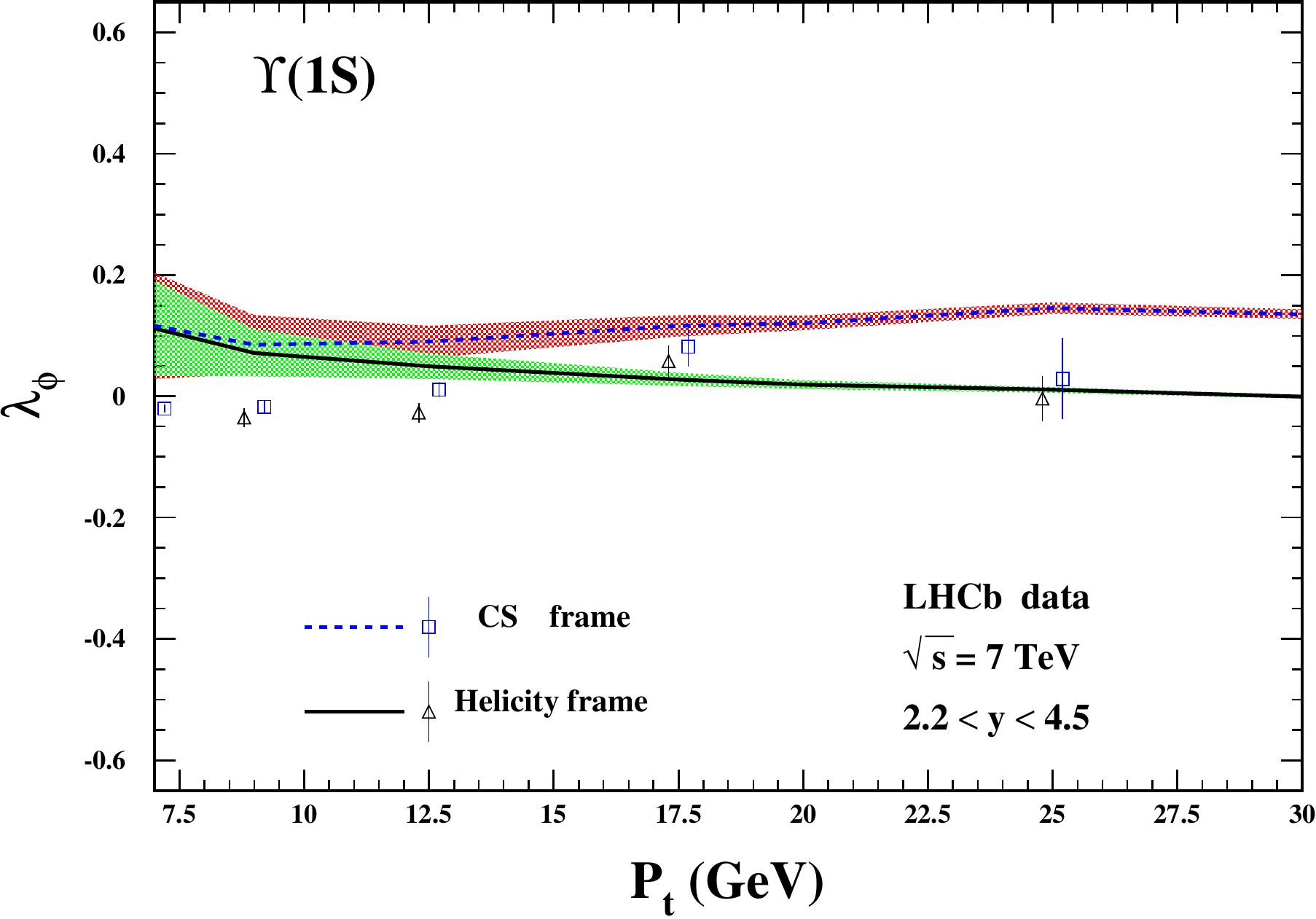}
  \includegraphics[width=5.0cm]{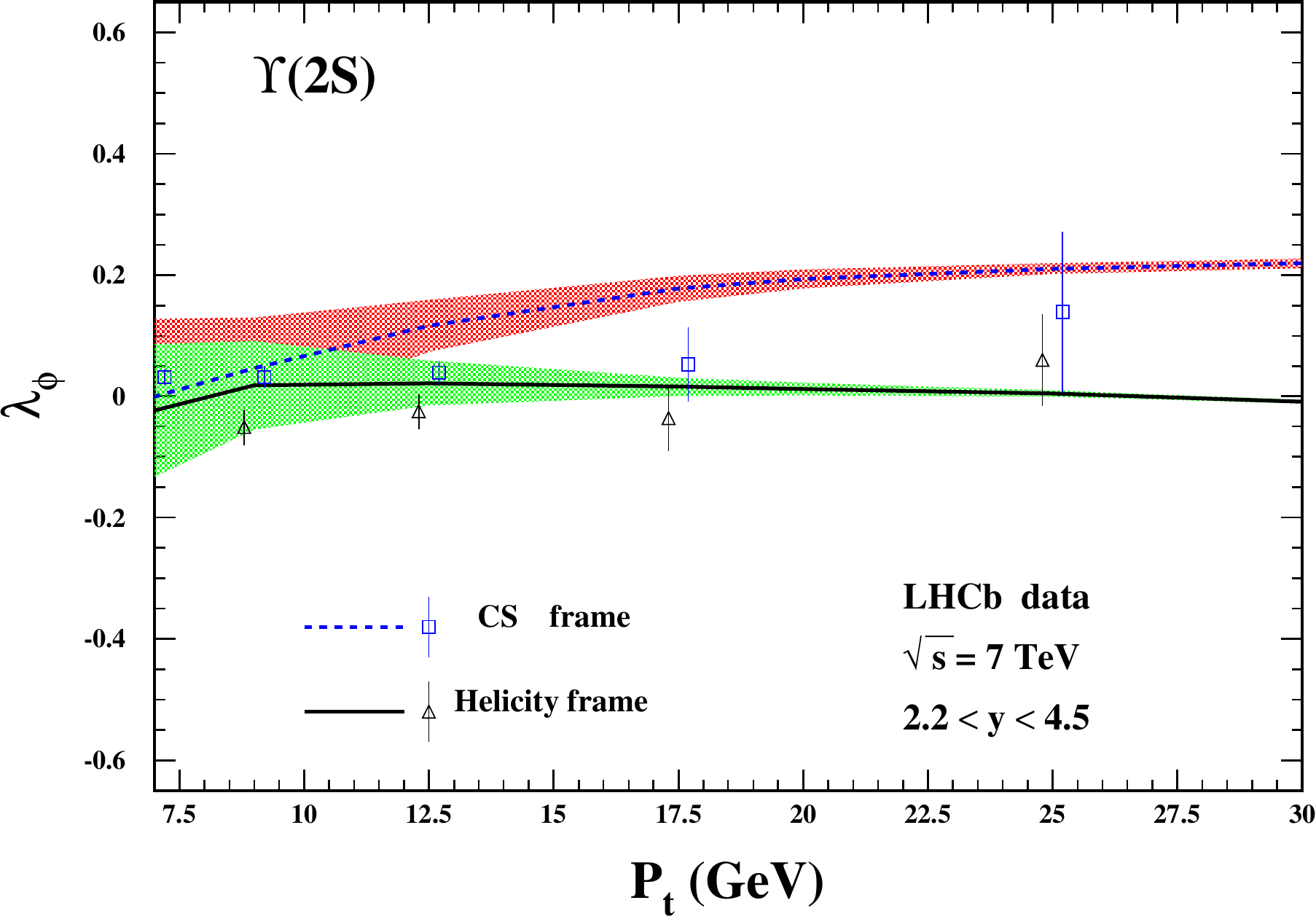}
  \includegraphics[width=5.0cm]{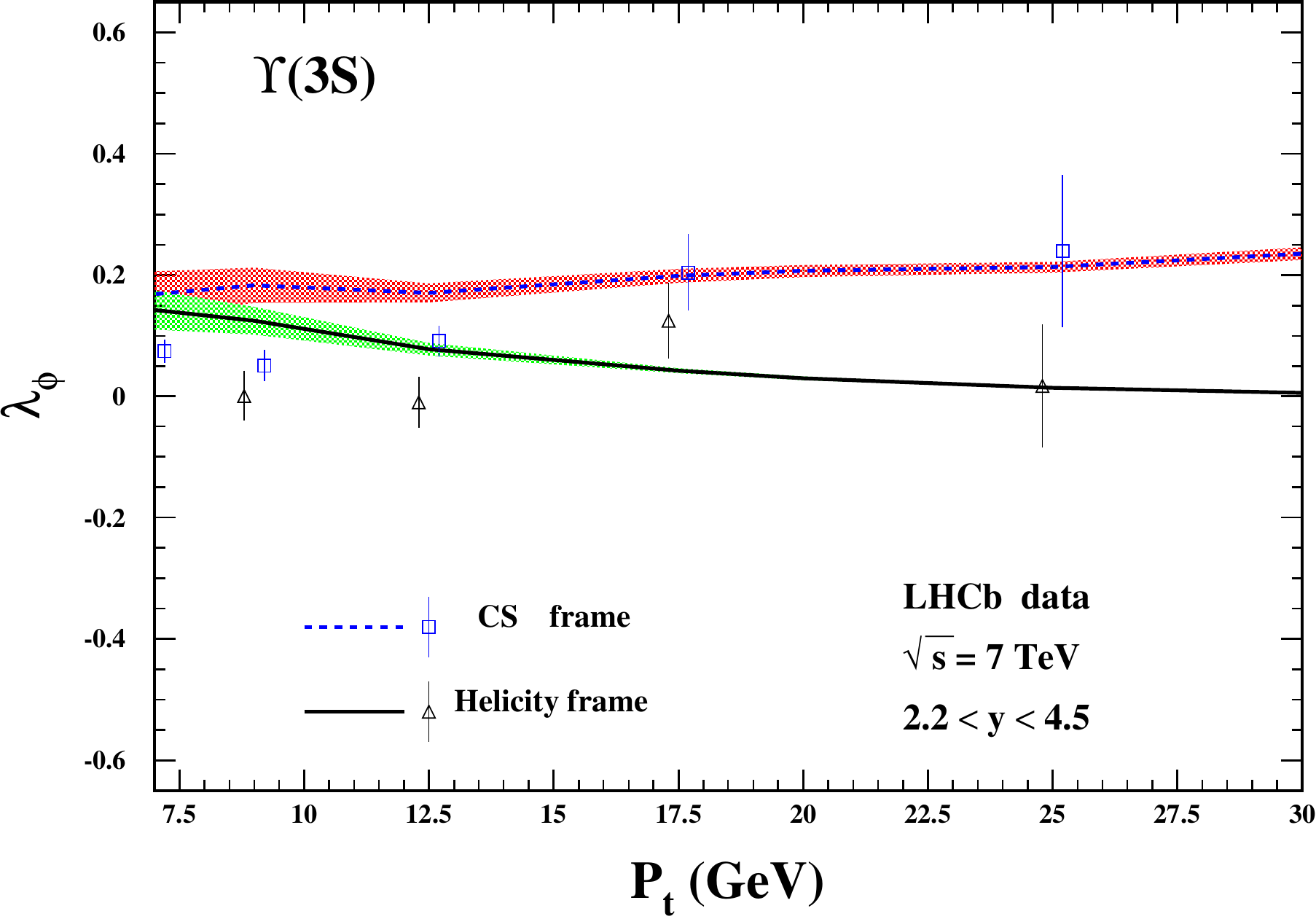} \\
  \caption{Polarization parameters $\lambda_{\theta}$(top), $\lambda_{\theta\phi}$(middle) and $\lambda_{\phi}$(bottom) for $\Upsilon$ hadroproduction in the forward rapidity region. From left to right:
  $\Upsilon(1S)$, $\Upsilon(2S)$ and $\Upsilon(3S)$. The LHCb data is from Ref.~\cite{Aaij:2017egv}.
   }
  \label{fig:pol:lhcb}
\end{figure*}

\begin{figure*} [htb]
  \centering
  \includegraphics[width=5.0cm]{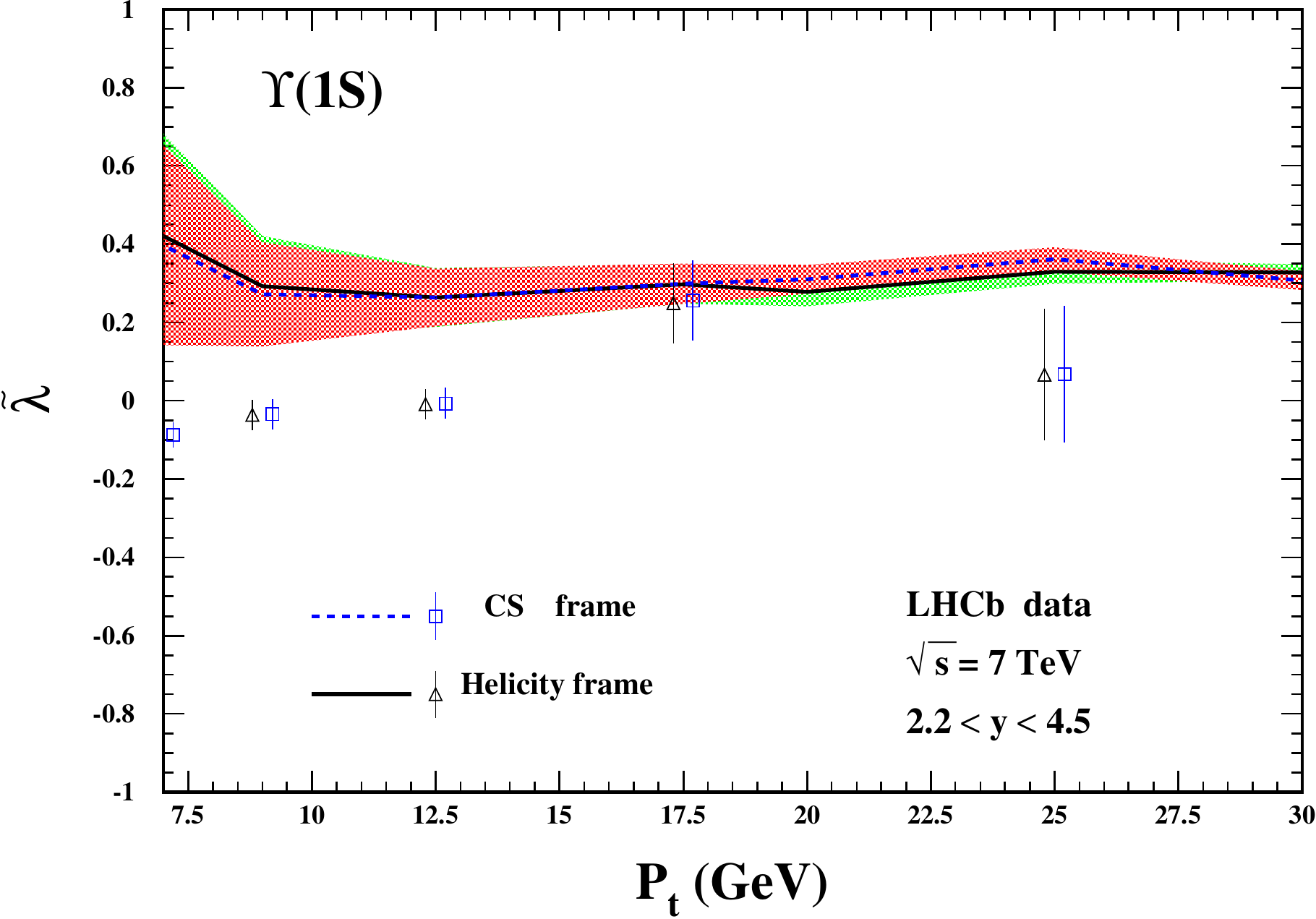}
  \includegraphics[width=5.0cm]{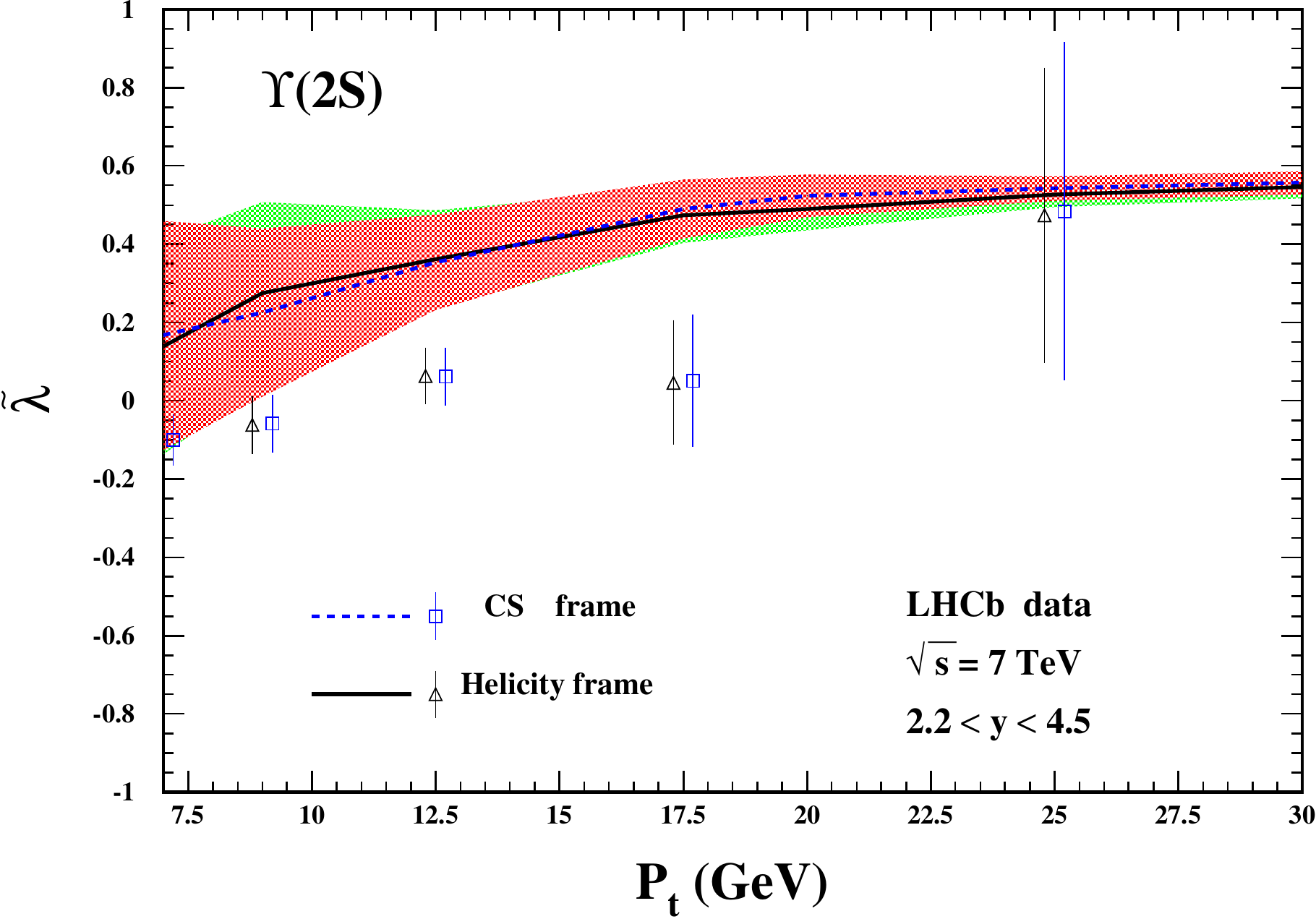}
  \includegraphics[width=5.0cm]{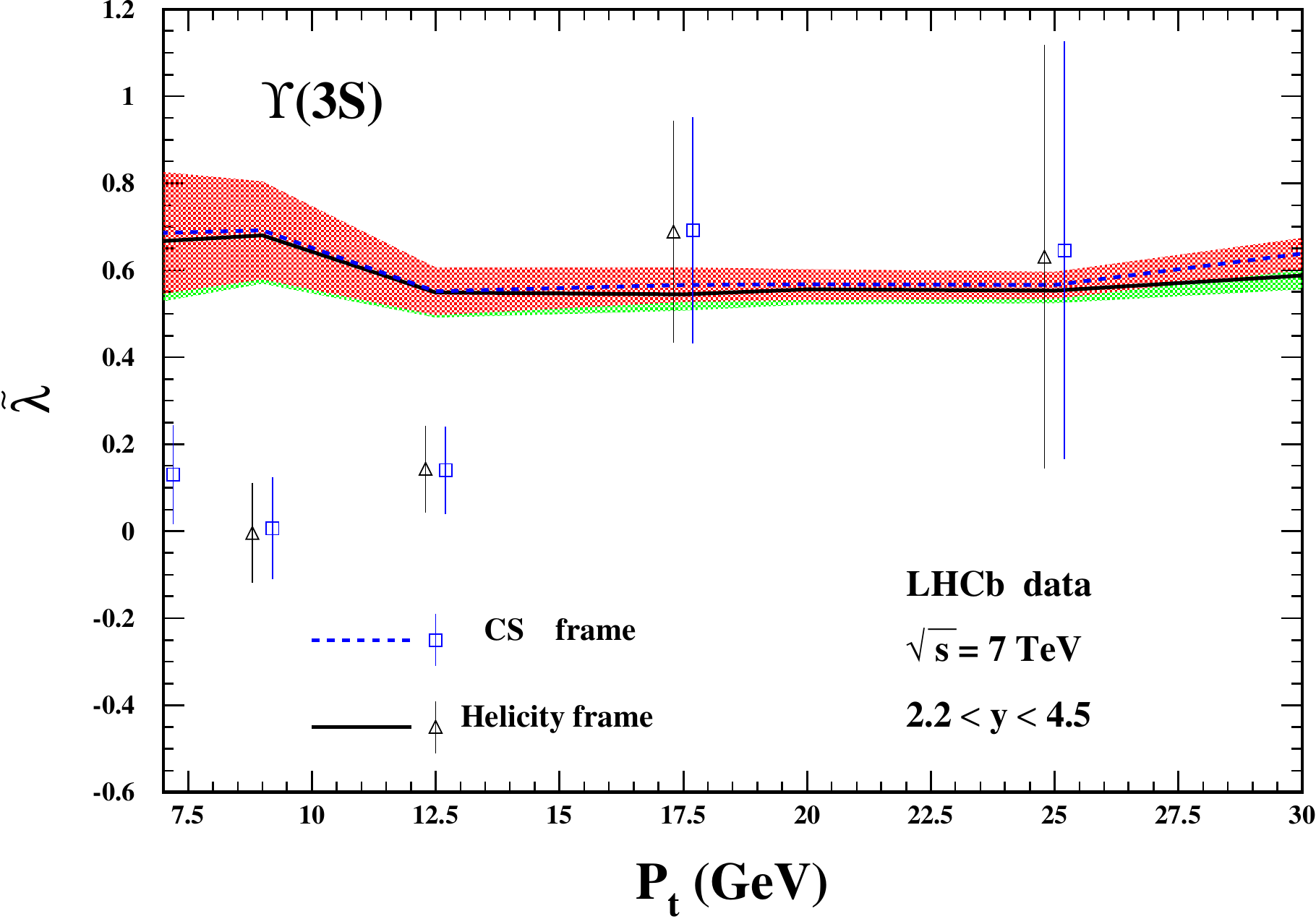}
  \caption{Polarization parameters $\widetilde{\lambda}$ for $\Upsilon$ hadroproduction in the forward rapidity region. From left to right:
  $\Upsilon(1S)$, $\Upsilon(2S)$ and $\Upsilon(3S)$. The LHCb data is from Ref.~\cite{Aaij:2017egv}.
   }
  \label{fig:lamTilde:lhcb}
\end{figure*}

The prediction of $\Upsilon$ polarization in the kinematic region relating to LHCb measurements are computed and presented in Figs.\ref{fig:pol:lhcb}
and \ref{fig:lamTilde:lhcb}.
For $\lambda_{\theta}$, our results provide a good description to the experimental measurements for $\Upsilon(1S,3S)$ in both the helicity and the CS frames.
For $\Upsilon(2S)$, the results are consistent with the experimental data in the CS frame,
while it is little higher than the data in the helicity frame.
Besides, the uncertainty of $\Upsilon(2S)$ is obviously larger at low $p_t$ region.
The discrepancy between theory and experiment at small $p_t$ is not surprising since the convergence of perturbative expansion is thought to be worse in this kinematical region and the data points with $p_t<8$ GeV were excluded in Ref.~\cite{Feng:2015wka} when extracting the LDMEs.

As regards $\lambda_{\theta\phi}$, our results provide a beautiful description to the data of $\Upsilon(3S)$
in both polarization frames.
For $\Upsilon(2S)$, the theory can cover the most measurements within the uncertainties.
This situation becomes worse for $\Upsilon(1S)$.
This situation becomes little bad for $\Upsilon(1S)$ since less data can be covered by the predictions.
Nevertheless, within 2$\sigma$ accuracy, all the measurements can be touched by the theory band.

For $\lambda_{\phi}$, our results are in (good, good, bad) agreement with the experimental data for ($\Upsilon(3S)$, $\Upsilon(2S)$ and $\Upsilon(1S)$), respectively.

For $\lambda_{\phi}$, our results are in good agreement with the available data for $\Upsilon(2S)$.
For $\Upsilon(1S)$ and $\Upsilon(3S)$, the predictions are little higher than the measurements in the low $p_t$ region, but
in the higher $p_t$ region ($p_t>15$GeV) the theory and experimental data are consistent.

In Fig.~\ref{fig:lamTilde:lhcb}, the frame-invariant quantity $\widetilde{\lambda}$ of $\Upsilon(nS)$ are compared with LHCb data~\cite{Aaij:2017egv}.
Again, the theoretical results are consistent between the two polarization frames
whereas
only one or two experimental
data points at higher $p_t$ region can be matched to theoretical predictions.
only part of experimental data points can be matched to theoretical predictions.

\subsection{The ratios of feed-down contributions}

\begin{figure*}[htb]
  \centering
  \includegraphics[width=5.0cm]{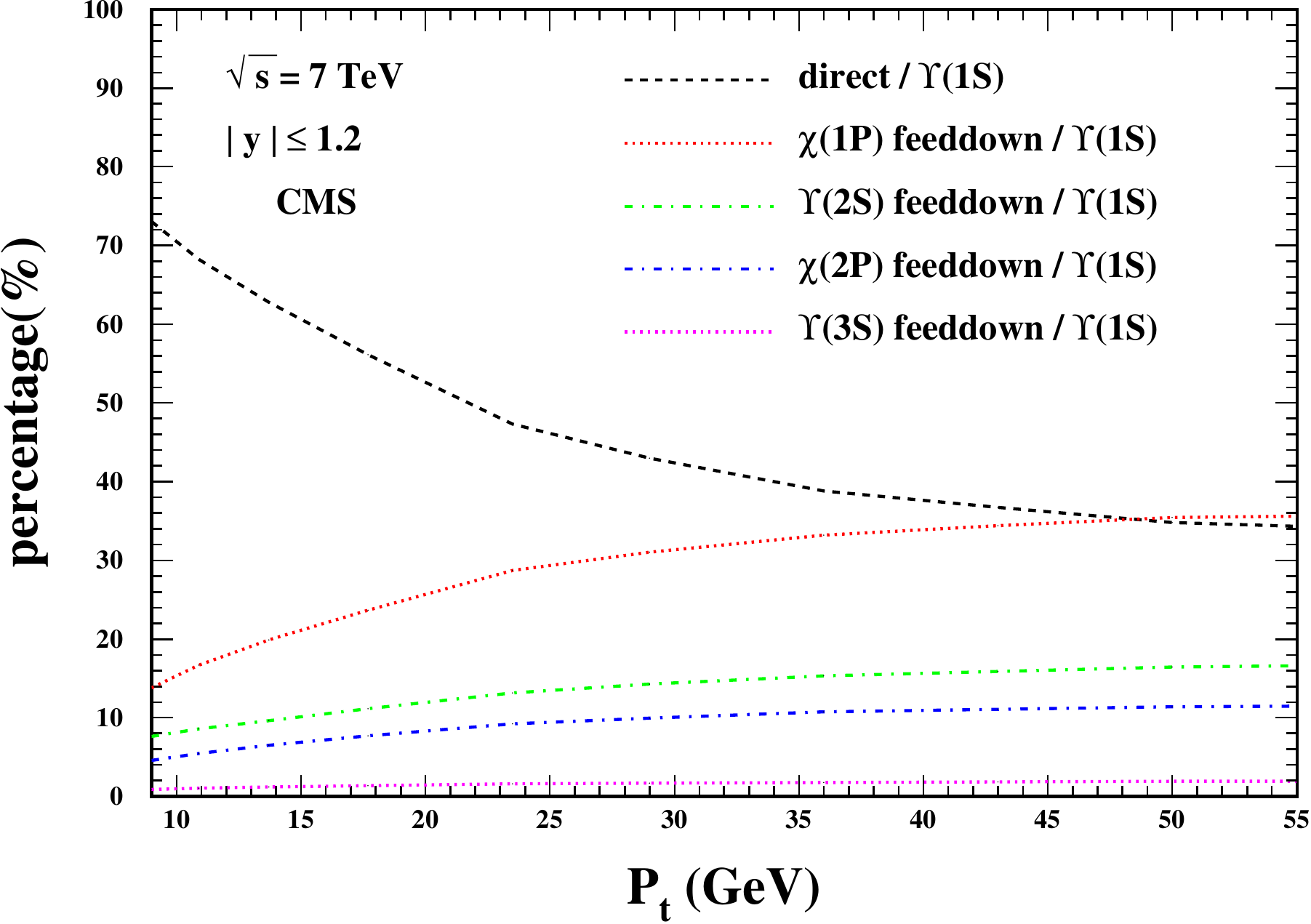}  \includegraphics[width=5.0cm]{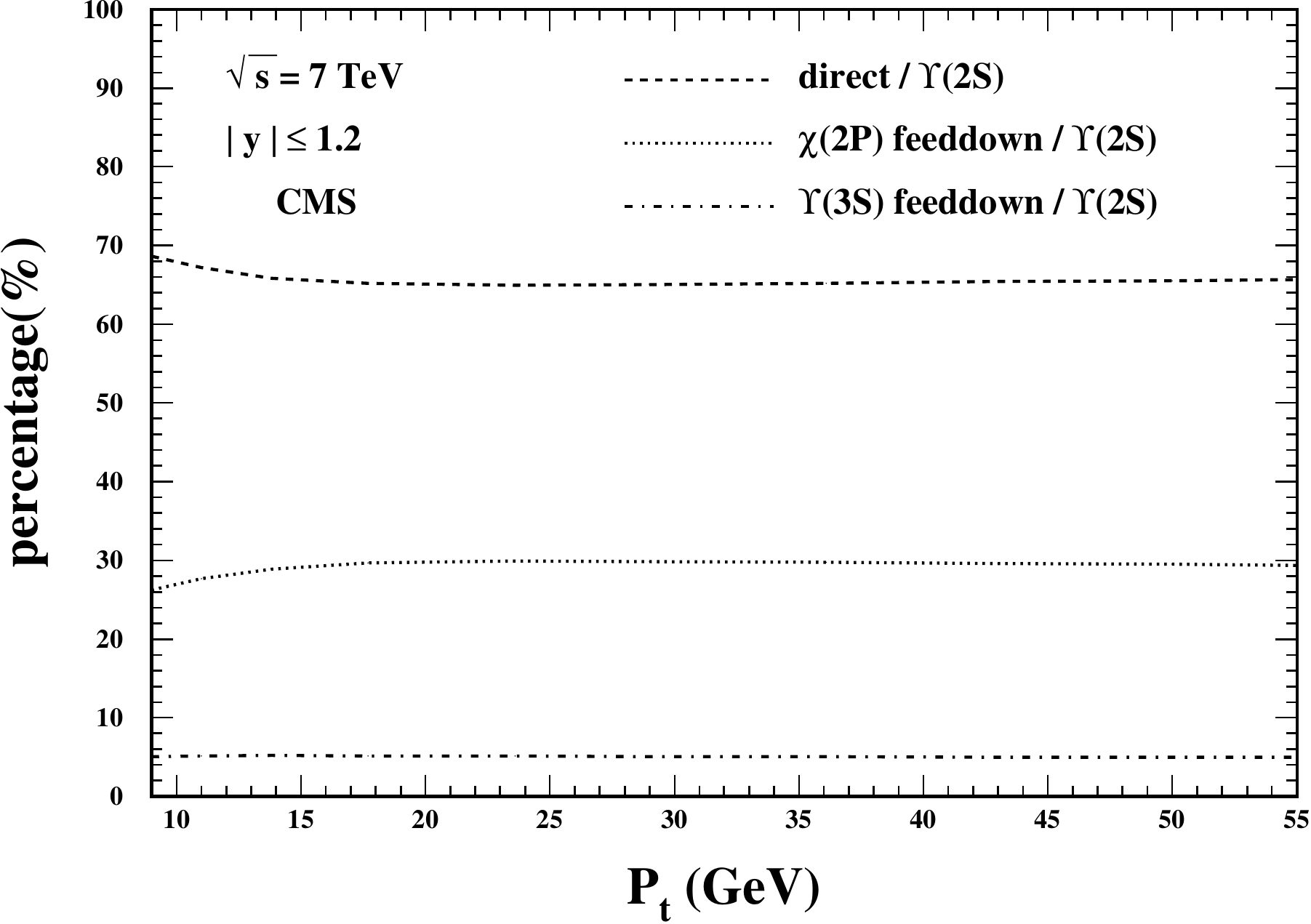} \includegraphics[width=5.0cm]{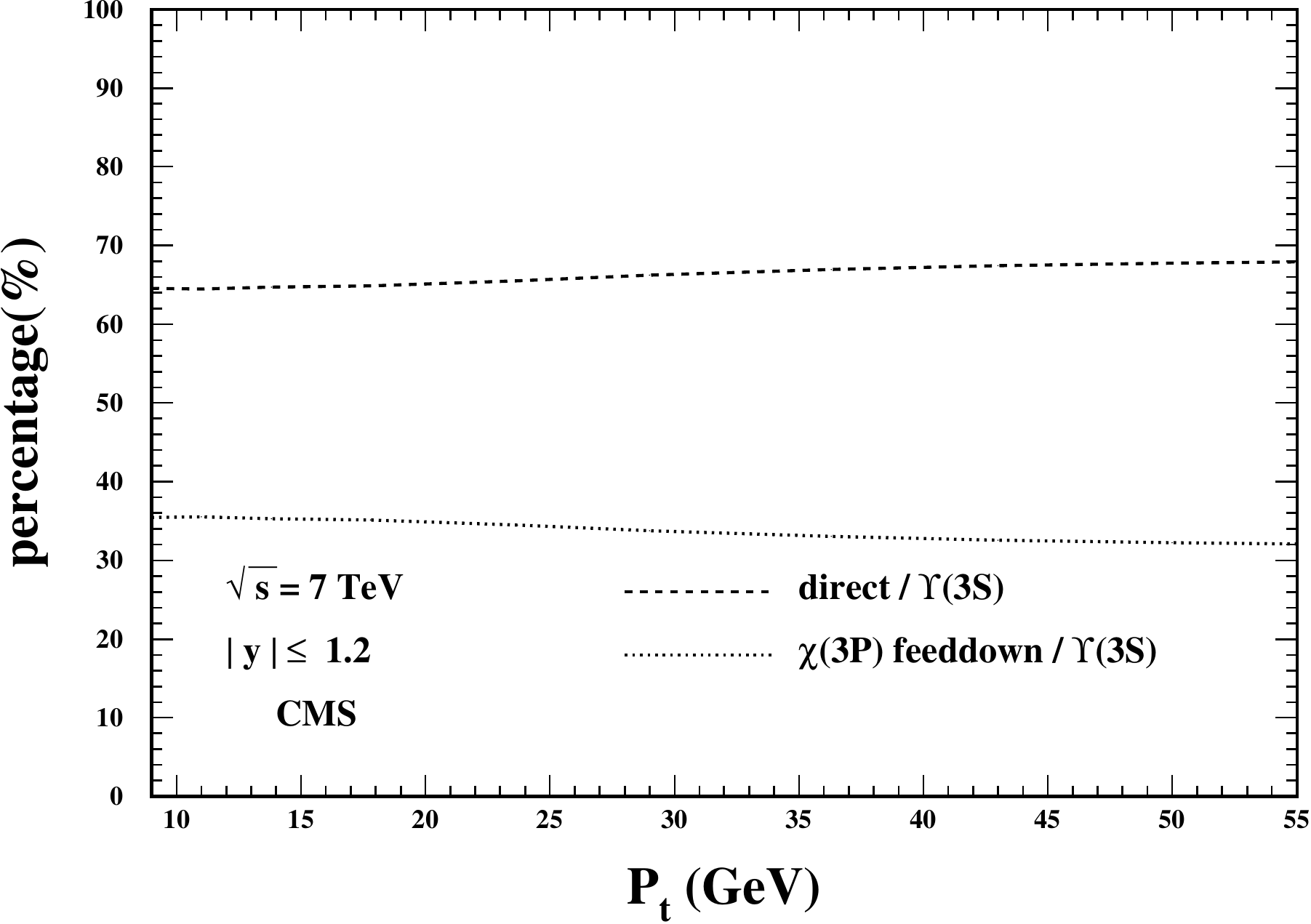} \\
  \includegraphics[width=5.0cm]{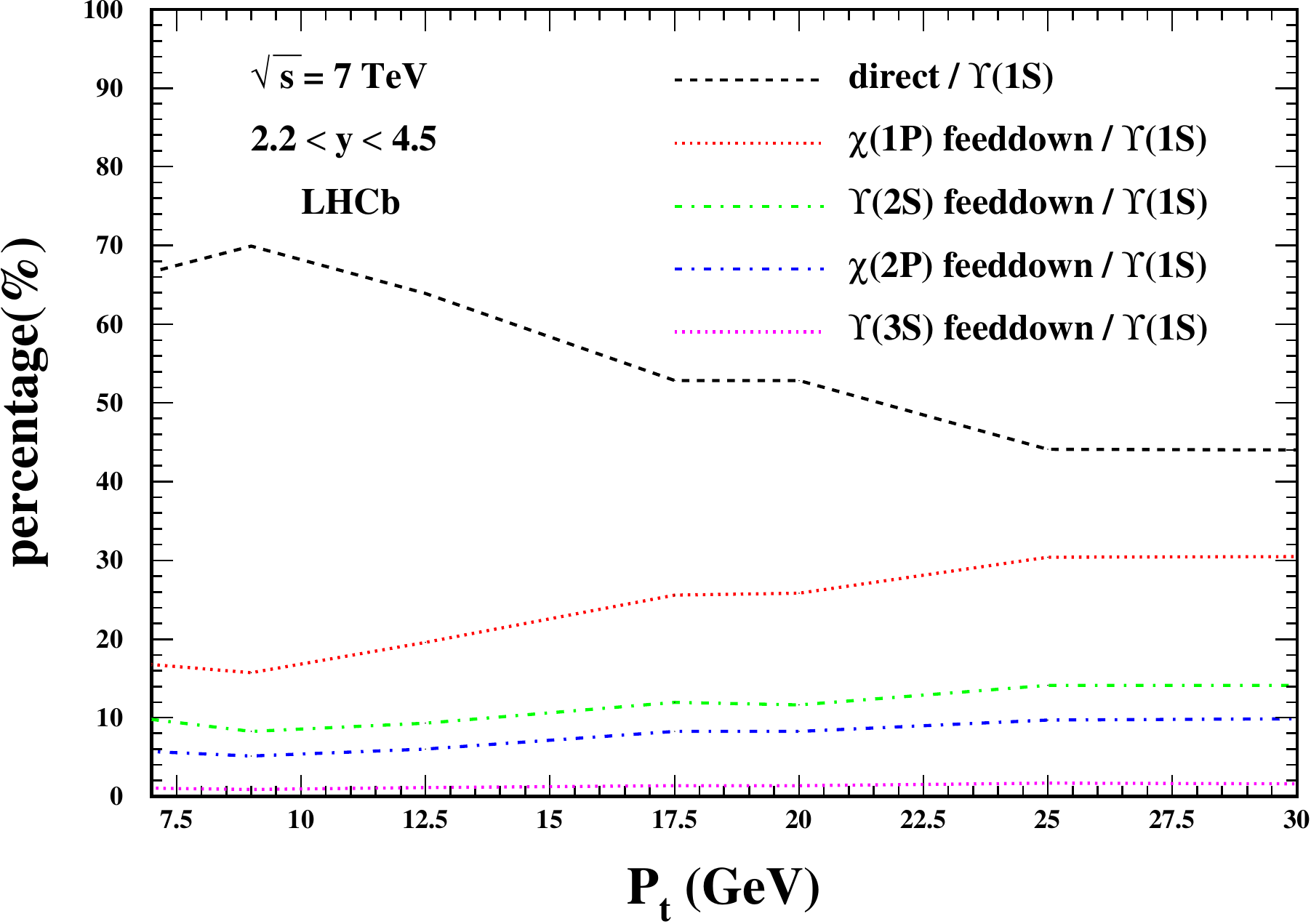}  \includegraphics[width=5.0cm]{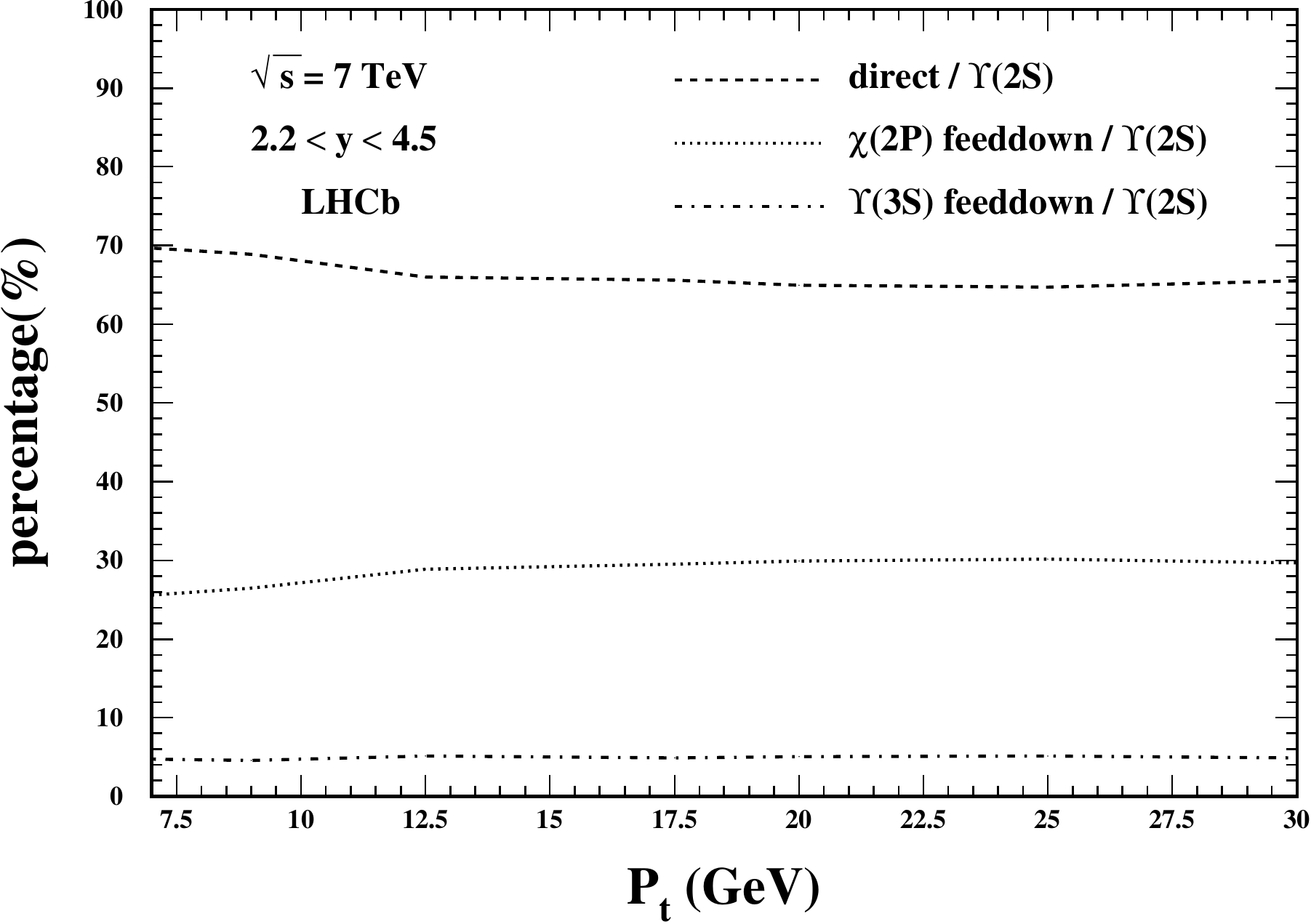} \includegraphics[width=5.0cm]{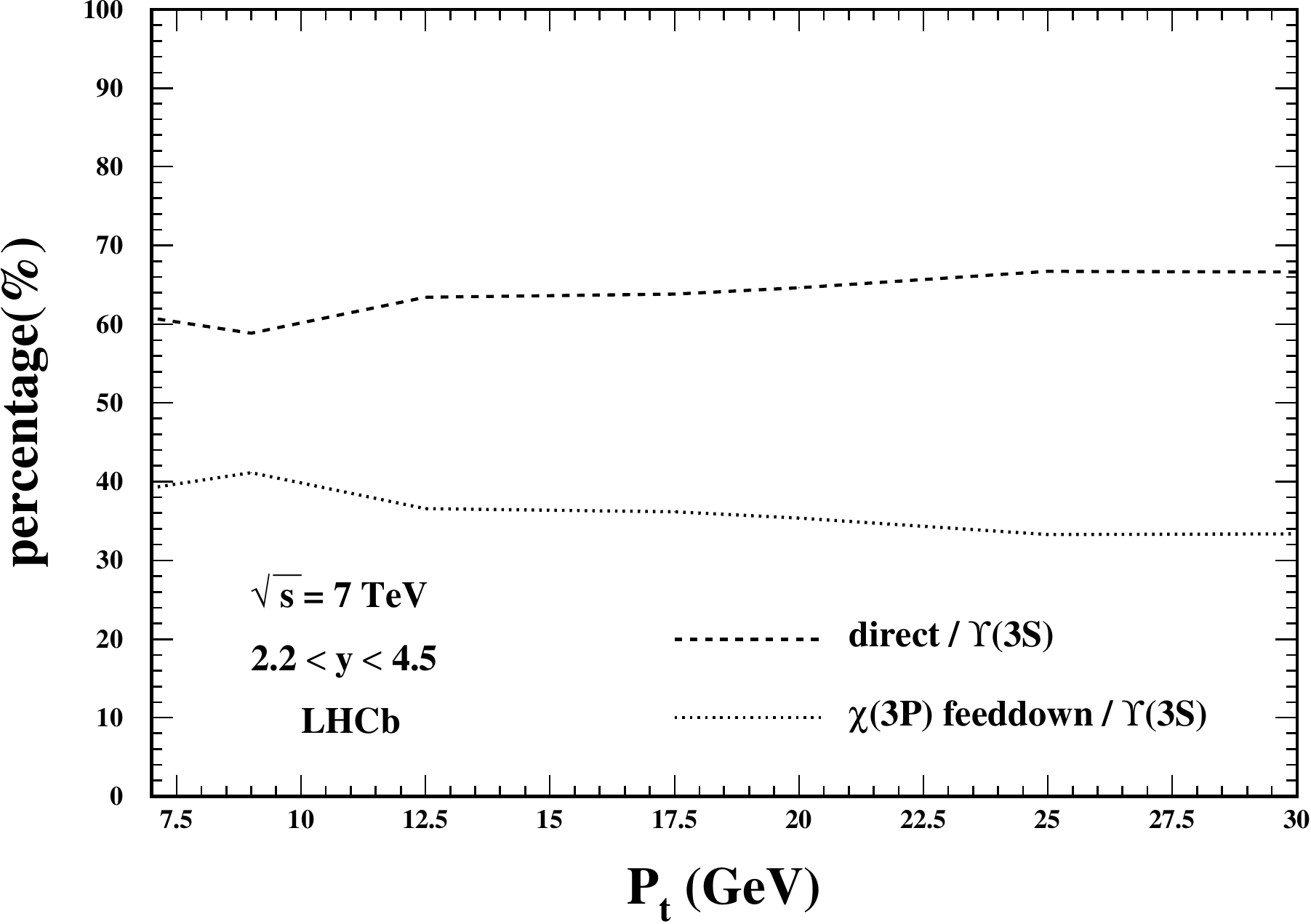} \\
  \caption{The ratio of contributions to the $\Upsilon(1S)$(left),$\Upsilon(2S)$(middle),$\Upsilon(3S)$(right) hadroproduction at the LHC.}
  \label{fig:ratio1}
\end{figure*}
\begin{figure*}[htb]
  \centering
  \includegraphics[width=5.0cm]{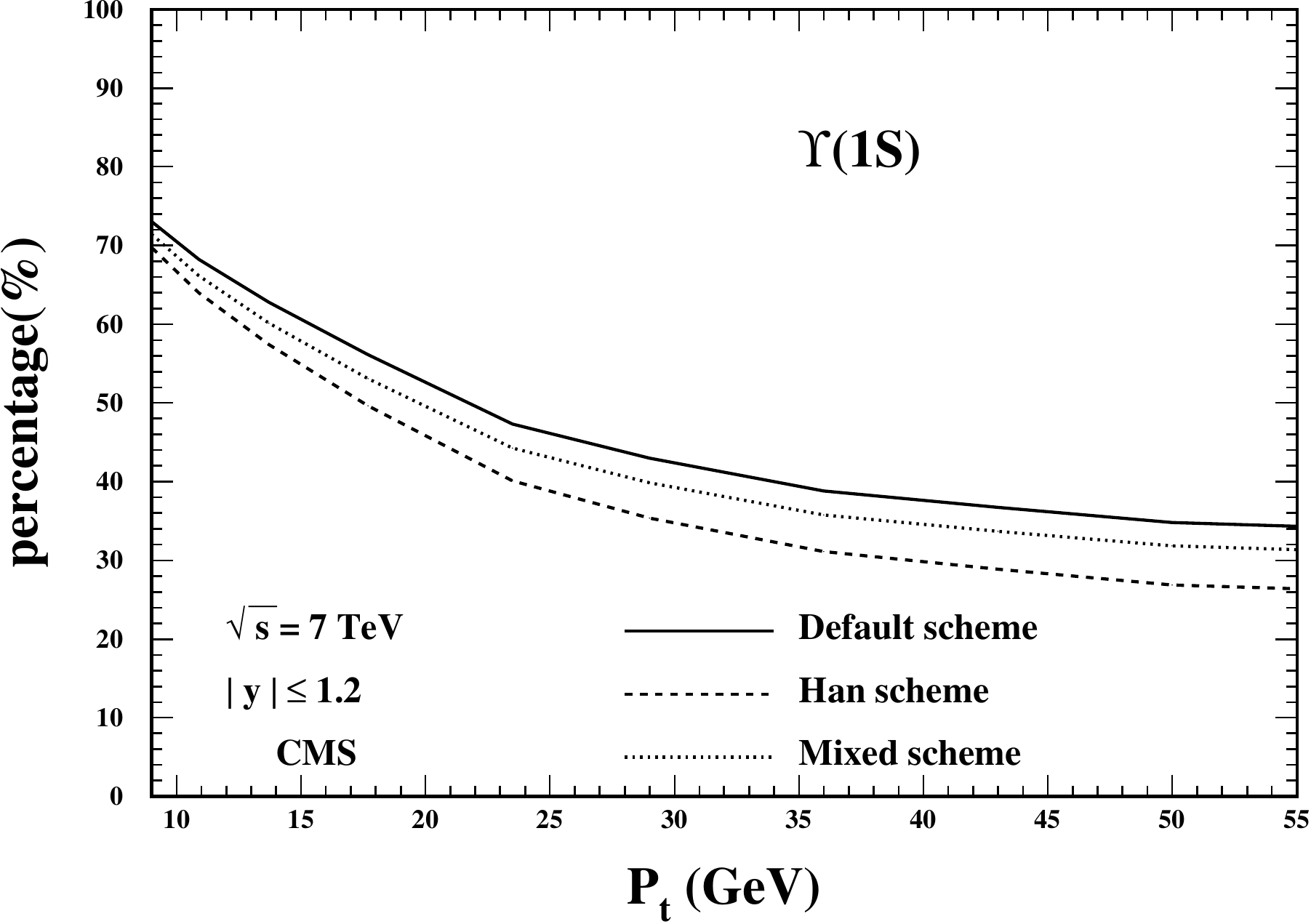}  \includegraphics[width=5.0cm]{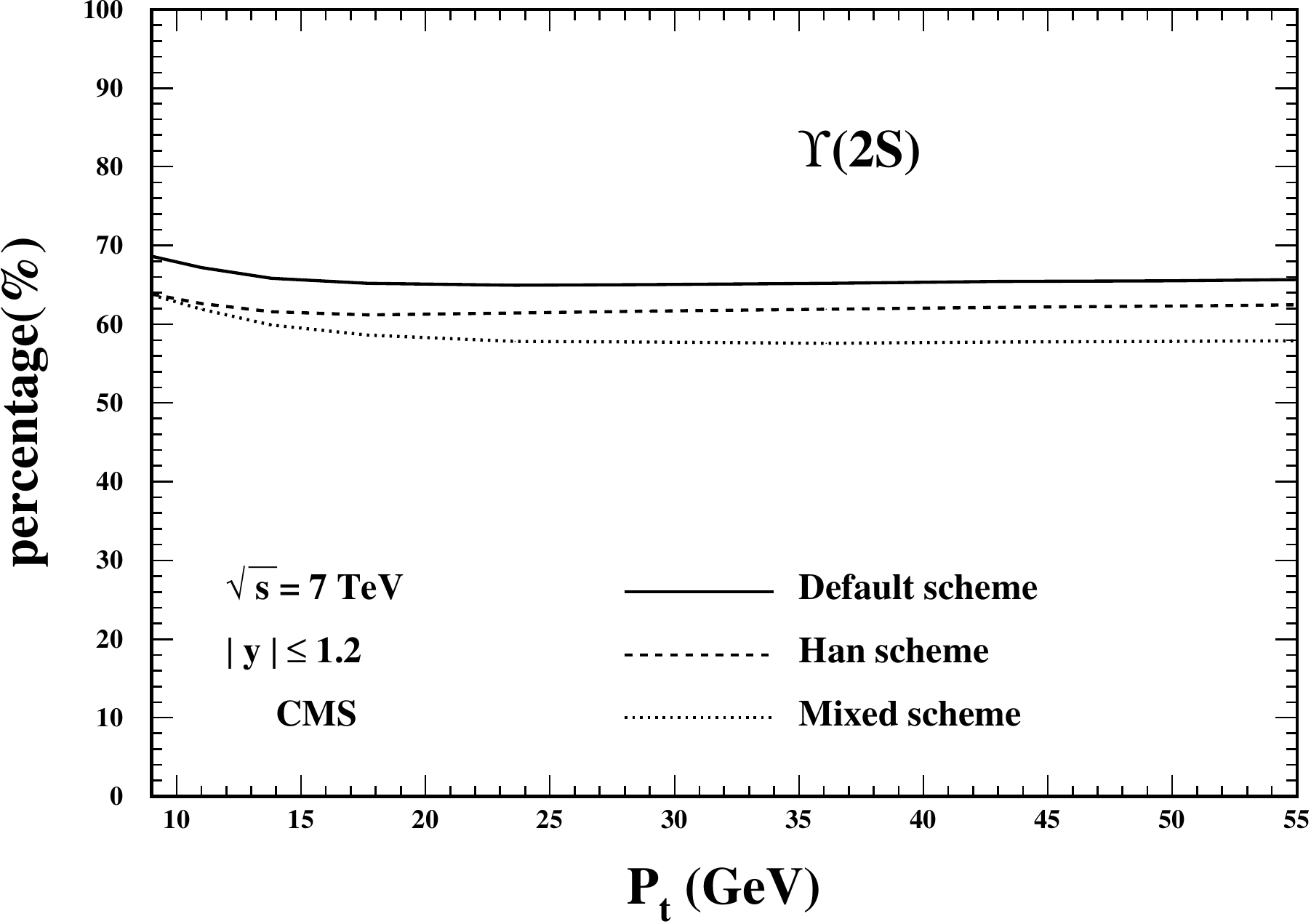} \includegraphics[width=5.0cm]{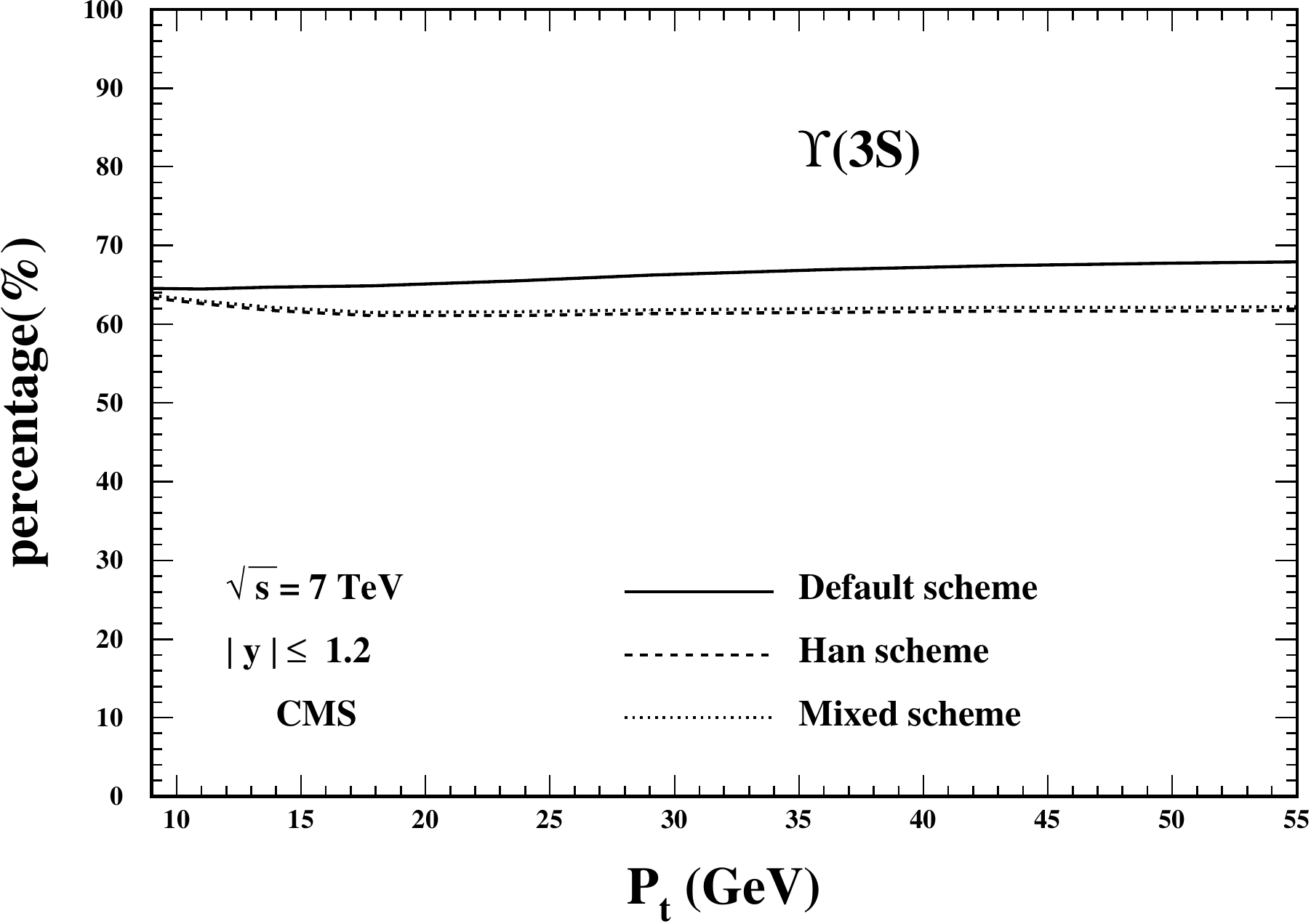} \\
  \includegraphics[width=5.0cm]{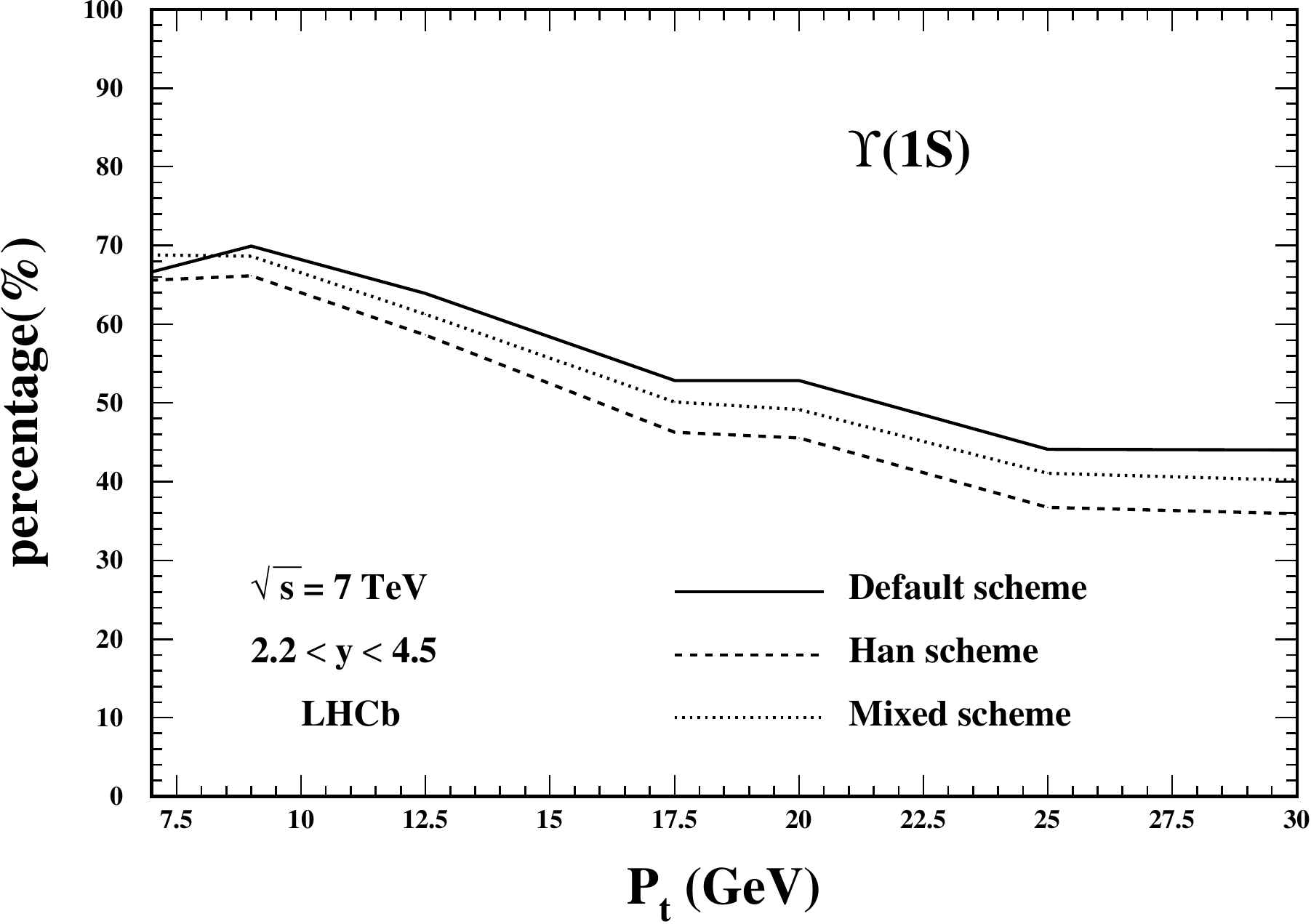}  \includegraphics[width=5.0cm]{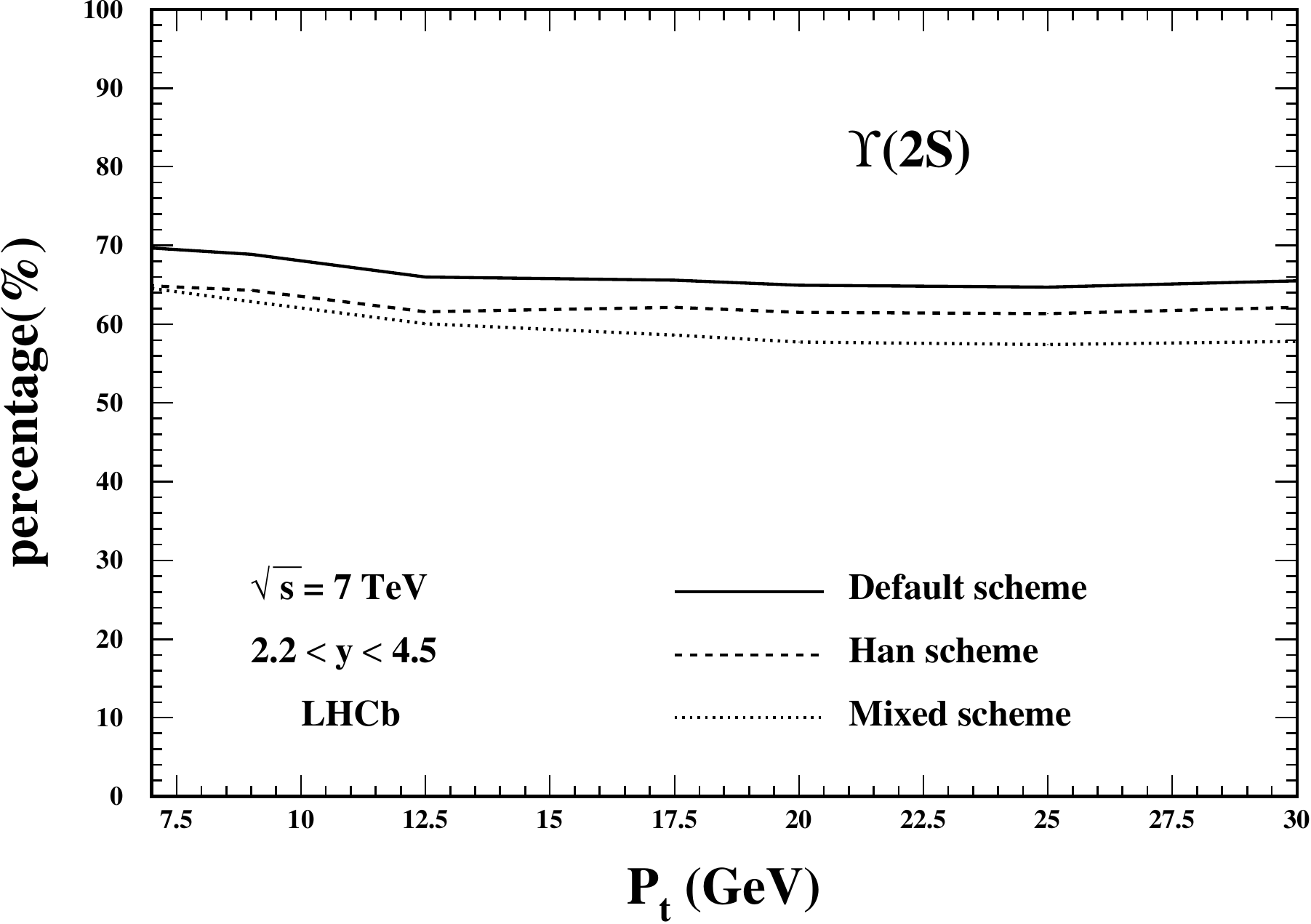} \includegraphics[width=5.0cm]{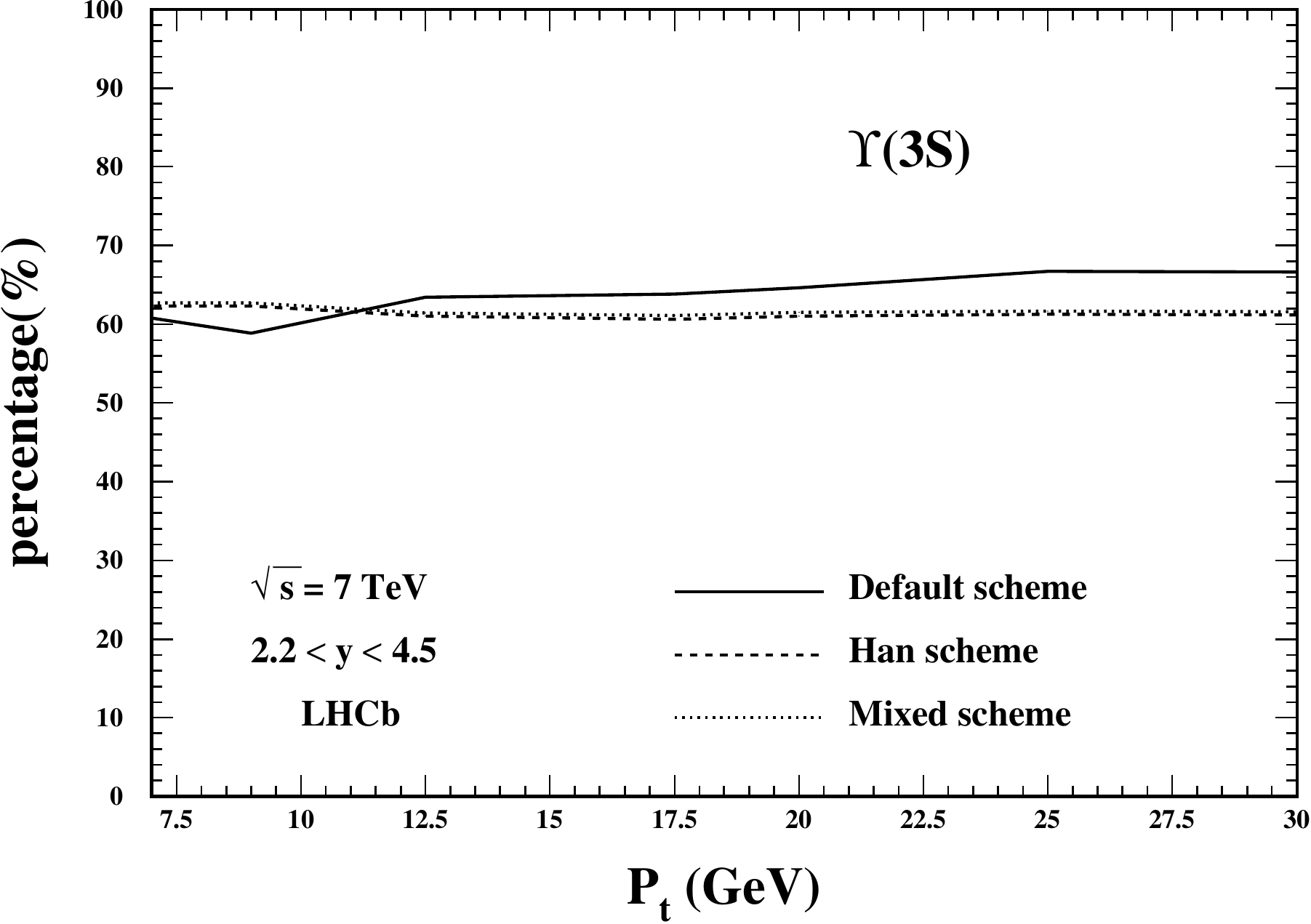} \\
  \caption{The ratio of the direct production over $\Upsilon(1S)$(left),$\Upsilon(2S)$(middle),$\Upsilon(3S)$(right) in three LDMEs schemes.}
  \label{fig:ratio2}
\end{figure*}

Here, we present the ratios of the feed-down contribution to $\Upsilon(1S,2S,3S)$ hadroproduction as a complement.
The ratios of all the feed-down channels are plotted in Fig.~\ref{fig:ratio1}, where we can find that the feed-down contributions are very important for the $\Upsilon$ production.
It can be seen that
in the $\Upsilon(3S)$ production, the feed-down from $\chi_{bJ}(3P)$ contributes more than 30\% in the whole $p_t$ region at CMS, while at LHCb, it can be even higher in low $p_t$ region.
For $\Upsilon(2S)$ production, the $\Upsilon(3S)$ feed-down contributes about 5\%, while the $\chi_{bJ}(2P)$ feed-down contribute about 30\% steadily in the whole $p_t$ region considered here.
For $\Upsilon(1S)$ production, the multiple feed-down contributions are presented in the plots,
where the feed-down contributions increase from 20\% to 60\% as transverse momentum $p_t$ becomes larger for
both LHCb and CMS windows.
The feed-down contribution from $\Upsilon(3S)$ to $\Upsilon(1S)$ ($\Upsilon(2S)$) production is less than 2\% (5\%), which seems to be negligible,
while the contributions from $\chi_{bJ}(nP)$ decay dominate the feed-down contributions to the corresponding $\Upsilon(nS)$ production.

To investigate the uncertainties of the ratios of the feed-down contributions
from the different sets of LDMEs, the other two LDMEs sets in Ref~\cite{Feng:2015wka} are used to compute the ratios.
To avoid confusion, we simply present the ratios of direct $\Upsilon$ production.
As comparison, the ratios via three LDMEs sets are presented in Fig.~\ref{fig:ratio2},
where the differences among the three curves are small for the $\Upsilon(1S,2S,3S)$ productions.

\section{\label{sec:summary}Summary and conclusion}

In this paper, a complete analysis on the $\Upsilon(1S,2S,3S)$ polarization is carried out.
All the three polarization parameters $\lambda_{\theta}$, $\lambda_{\theta\phi}$, $\lambda_{\phi}$ for $\Upsilon(1S,2S,3S)$ hadroproduction have been calculated at QCD NLO within NRQCD framework.
The frame-invariant quantity $\widetilde{\lambda}$ at CMS and LHCb are also investigated.
As a complement, we present the ratios of the feed-down contributions to $\Upsilon(1S,2S,3S)$ hadroproduction.

Before comparing our results with experimental data, it is important to mention that $\lambda_{\theta\phi}$ in the helicity frame,
which is already investigated in Ref.~\cite{Feng:2018ukp}, should be exactly zero in symmetric rapidity region.
Although most of data from CMS are consistent with this within 1$\sigma$ level, there do exist several data points which needs 2$\sigma$,
and there even exists one point in the case of $0.6\le |y| \le 1.2$ for $\Upsilon(3S)$ which  is outside $2\sigma$ range.
Therefore, We suggest that in order to improve the experimental measurement at CMS (in symmetric rapidity region), $\lambda_{\theta\phi}$ in the
helicity frame should be constrained to zero.
Due to this situation in the experimental measurements, in our comparison, ``very good'', ``good'', ``acceptable'' and ``bad'' are used
if theoretical results and experimental data are consistent with each other at $1\sigma$, $2\sigma$, $3\sigma$ and $>3\sigma$ levels, respectively.

For $\lambda_\theta$ and $\lambda_{\theta\phi}$, our results can describe CMS data quite well in both the helicity and CS frames, as shown in Figs~\ref{fig:pol:cms06} and \ref{fig:pol:cms12}.
However for LHCb data, although most data can still be well described, there do exist some points which are inconsistent with theoretical predictions within $3\sigma$ level.
Things become worse for $\lambda_\phi$. Among all the data points from CMS and LHCb in both the helicity and CS frames, it is found that about $1/3$ of them can be described within 1$\sigma$ level, and another $1/4$ of them are within 2$\sigma$ level, while about $1/6$ of them are inconsistent with theoretical predictions within 3$\sigma$ level.
Due to this, for the frame independent parameter $\tilde{\lambda}$, only about 60\% of experimental data can be described by theoretical predictions within 2$\sigma$ level, although
their results consist quite well with themselves in the two frames .


In addition, the ratios of the feed-down contributions to $\Upsilon(1S,2S,3S)$ hadroproduction
are presented with different LDMEs schemes. The results indicate the feed-down contributes
more than 30\% to $\Upsilon(1S,2S,3S)$ hadroproduction, which emphasizing the importance of the feed-down contributions.

\acknowledgments{
The work were achieved by using the HPC Cluster of ITP-CAS.
This work was supported in part by the National Natural Science Foundation of
China with Grants Nos.
11905292,
11535002, 11675239,
11745006,
11821505, 11947302 and 11975242.
It was also supported by Key Research Program of Frontier Sciences, CAS, Grant No. QYZDY-SSW-SYS006 and Y7292610K1.
Y.F. would like to thank CAS Key Laboratory of Theoretical Physics, Institute of Theoretical Physics (ITP), CAS, for the very kind invitation and hospitality.

}

\bibliography{nrqcd}

\end{document}